%% file: 12892-Final.tex
\documentclass{fundam}

\usepackage[numbers,sort&compress]{natbib}
\usepackage{graphicx,lineno,hyperref,pgfkeys}
\newcommand{\aget}[2]{{#1}[{#2}]}
\newcommand{\anc}[2]{\aread{{#1}^*}{#2}}

\newcommand{\aread}[2]{{#1}[{#2}]}
\newcommand{\aset}[3]{{#1}[{#2}] := {#3}}
\newcommand{\awrite}[3]{{#1}[{#2}\mapsto{#3}]}
\newcommand{\astop}{{}^*}
\newcommand{\cnv}[1]{{#1}^{\sf T}}
\newcommand{\cnvop}{\cnv{}}
\newcommand{\comp}{\cdot}
\newcommand{\cpl}[1]{\overline{#1}}
\newcommand{\cplop}{\cpl{\rule{0em}{1.2ex}~\,}}
\newcommand{\down}[1]{${#1}{\downarrow}$}
\newcommand{\fc}{\mathrm{fc}}
\renewcommand{\iff}{\mathrel{\Leftrightarrow}}
\renewcommand{\implies}{\mathrel{\Rightarrow}}
\newcommand{\ind}{\hspace*{1em}}
\newcommand{\join}{\sqcup}
\newcommand{\kroot}[2]{\mbox{root ${#1}$ ${#2}$}}
\newcommand{\kroots}[1]{\mbox{roots ${#1}$}}
\newcommand{\lbot}{\bot}
\newcommand{\lleq}{\sqsubseteq}
\newcommand{\ltop}{\top}
\newcommand{\map}{\mathrm{map}}
\newcommand{\matt}[9]{{\left(\!\begin{array}{c@{~~}c@{~~}c} #1 & #2 & #3 \\ #4 & #5 & #6 \\ #7 & #8 & #9 \end{array}\!\right)}}
\newcommand{\maxi}{\mathsf{M}}
\newcommand{\meet}{\sqcap}
\newcommand{\opzero}{0}
\newcommand{\opsucc}{\mathit{succ}}
\newcommand{\plusop}{{}^+}
\newcommand{\psuc}{\mathsf{S'}}
\newcommand{\suc}{\mathsf{S}}
\newcommand{\rank}{\mathit{rank}}
\newcommand{\unit}{1}
\newcommand{\wcc}{\mathrm{wcc}}
\newcommand{\xs}{\mathit{xs}}
\newcommand{\zero}{\mathsf{Z}}

\newtheorem{thm}{Theorem}
\newenvironment{prf}{\trivlist\PRstyle\item[]{\bfseries Proof:}}{\endtrivlist}
\newenvironment{prfof}[1][]{\trivlist\PRstyle\item[]{\bfseries Proof #1:}\newline}{\QED\endtrivlist}
\newenvironment{prfofnoqed}[1][]{\trivlist\PRstyle\item[]{\bfseries Proof #1:}\newline}{\endtrivlist}

\newcommand{\git}[1]{https://foss.heptapod.net/isa-afp/afp-devel/-/#1/9221bd1bd835775474be509404e45e65f26ee00d/thys/Relational_Disjoint_Set_Forests}
\newcommand{\tree}[1]{\href{\git{tree}}{#1}}
\newcommand{\blob}[2]{\href{\git{blob}/#1}{#2}}
\newcommand{\afplogo}{\includegraphics[height=2ex]{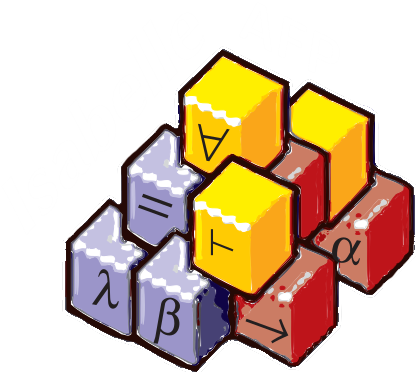}}
\newcommand{\afp}[1]{~\blob{#1}{\afplogo}}
\newcommand{\afplemma}[1]{\afp{\pgfkeysvalueof{/line numbers/lemma #1}}}
\newcommand{\afptheorem}[1]{\afp{\pgfkeysvalueof{/line numbers/theorem #1}}}

\input{afp-dict}

\begin{document}


\setcounter{page}{77}
\publyear{24}
\papernumber{2189}
\volume{192}
\issue{1}

\finalVersionForARXIV

\title{Relation-Algebraic Verification of Disjoint-Set Forests}

\author{Walter Guttmann\thanks{Address for correspondence: Department of Computer Science and Software Engineering,
                                          University of Canterbury, Christchurch, New Zealand.  \newline \newline
                    \vspace*{-6mm}{\scriptsize{Received January 2023; \ accepted July 2024.}}}
                \\
        Department of Computer Science and Software Engineering \\
        University of Canterbury \\
        Christchurch, New Zealand \\
       walter.guttmann@canterbury.ac.nz
       }

\runninghead{W. Guttmann}{Relation-Algebraic Verification of Disjoint-Set Forests}

\maketitle

\begin{abstract}
  This paper studies how to use relation algebras, which are useful for high-level specification and verification, for proving the correctness of lower-level array-based implementations of algorithms.
  We give a simple relation-algebraic semantics of read and write operations on associative arrays.
  The array operations seamlessly integrate with assignments in computation models supporting while-programs.
  As a result, relation algebras can be used for verifying programs with associative arrays.
  We verify the correctness of an array-based implementation of disjoint-set forests using the union-by-rank strategy and find operations with path compression, path splitting and path halving.
  All results are formally proved in Isabelle/HOL.
  This paper is an extended version of \cite{Guttmann2020a}.
\end{abstract}


\section{Introduction}
\label{section.introduction}

Relations, relation algebras, Kleene algebras and similar structures have been used for various aspects of program semantics, in particular, to model control flow, refinement and data structures \cite{BackWright1998,HoareHe1998,Maddux1996,SchmidtStroehlein1989}.
For example, the control-flow of while-programs can be modelled in Kleene algebras with tests, where the Kleene star is used to define the semantics of while-loops \cite{BackWright1999,Kozen1997}.
Program transformations and refinements can be carried out algebraically; for example, see \cite{BackWright1999,Kozen1994}.
On the data side, relations are intimately connected with graphs through their adjacency matrices, whence the data-flow of graph algorithms can be modelled using relation algebras, frequently extended by a Kleene star to describe transitive closure or reachability in graphs \cite{BackhouseCarre1975,Berghammer1999,BerghammerKargerWolf1998,BerghammerStruth2010,GondranMinoux2008,HoefnerMoeller2012,Moeller1993}.
Relations as a generalisation of functions are also useful for the specification and derivation of functional programs \cite{BirdMoor1997}.

This paper extends the relation-algebraic method by verifying algorithms which use associative arrays as data.
Hoare logic \cite{Hoare1969} is commonly used for verifying programs.
A verification condition generator automatically derives from the structure of the program a collection of logical statements whose proof implies correctness of the program.
When applied to graph algorithms using a relation algebra to represent graphs, the verification conditions are simply relation-algebraic formulas.
They can be discharged by a combination of manual and automated reasoning in relation algebras \cite{BerghammerStruth2010}.

When modelling graphs, the operations of relation algebras work on entire relations.
This abstract view is useful for specification and verification, but typically not intended directly for implementation.
In contrast, efficient algorithms are often expressed at a lower level, in particular, using arrays.
For example, the pseudo-code for disjoint-set forests in \cite{CormenLeisersonRivest1990} uses two arrays: one for the rank of a node and one for its parent.
A difference between these arrays is their type: the rank of a node is a natural number while the parent of a node is also a node.

This paper addresses the gap between the benefits of relation algebras for high-level specification and verification on the one hand, and the need for lower-level formalisms to express efficient algorithms on the other hand.
Specifically, we show how relation algebras, Kleene algebras and related structures can be used for proving the correctness of array-based implementations of algorithms.

An associative array is just a finite mapping from a set of indices to a set of values, hence a relation.
The term `array' often implies that the set of indices is an interval of integer numbers, but it can be an arbitrary finite set for associative arrays.
The rank array of a disjoint-set forest maps nodes to natural numbers, making it a heterogeneous relation.
The parent array maps nodes to nodes, which gives a homogeneous relation.

In the present paper, we focus on associative arrays with the same index and value sets, but this is not a fundamental limitation.
Without assuming any specific structure on the index/value set, we give a simple relation-algebraic semantics of reading from and writing to an array.
These access operations can occur in assignments in while-programs, and are therefore amenable to the usual verification techniques.
The generated verification conditions are relation-algebraic formulas using the semantics of the array operations.

As a case study, we implement disjoint-set forests in a way that is close to the pseudo-code in \cite{CormenLeisersonRivest1990} and verify their correctness in Isabelle/HOL \cite{NipkowPaulsonWenzel2002}.
This facilitates the use of relation-algebraic reasoning about algorithms expressed at a low level.
In this paper we do not go below the pseudo-code level of \cite{CormenLeisersonRivest1990}, for example, to a concrete programming language.
That said, associative arrays are commonly used to implement disjoint-set forests, not just in pseudo-code.
For example, the JGraphT library for Java (version 1.5.1, see \url{https://jgrapht.org/}) uses finite maps, and the Boost library for C++ (version 1.79.0, see \url{https://www.boost.org/}) uses finite maps from elements to integer indices and vectors (integer-indexed arrays whose size can change) for the actual data structure.
We formalise specifications and implementations and carry out the verification using relation algebras, Kleene algebras and related structures rather than Isabelle/HOL's finite mappings and relations in order to show that algebraic methods can be applied for lower-level reasoning.

\medskip
The contributions of this paper are:
\begin{itemize}
\item A simple relation-algebraic semantics of selective read and write in associative arrays.
\item Verification of the correctness of disjoint-set forests in Kleene relation algebras.
      This includes path compression, path splitting, path halving and union-by-rank.
      Additional structure is assumed to deal with ranks, which are natural numbers.
\item Constructive proof of a theorem of Kleene relation algebras (Theorem \ref{theorem.root}.\ref{theorem.root.point}) using an imperative program.
\item Formalisation of the above and all other results in Isabelle/HOL.
\end{itemize}
Proofs of all results can be found in the Isabelle/HOL theory files available in the Archive of Formal Proofs \cite{Guttmann2020c}.
The PDF version of this paper includes links to the respective theorems; click on occurrences of \tree{\afplogo} to follow these links.
Appendix \ref{section.proofs} contains proofs for a sample of theorems.

In Section \ref{section.motivation} we motivate the relation-algebraic approach.
Section \ref{section.relations} introduces the algebraic framework for the remainder of this paper including relation algebras and Kleene algebras.
We give a simple semantics of read and write access to associative arrays in Section \ref{section.arrays} and discuss basic properties.
The semantics of disjoint-set forests is provided in Section \ref{section.disjoint-sets}.
Forming the main part of this paper, Sections \ref{section.verification-init}--\ref{section.verification-union} describe our Isabelle/HOL verification of the total correctness of key operations on disjoint-set forests.
Section \ref{section.verification-init} deals with initialising disjoint-set forests by applying the make-set operation to all elements.
Section \ref{section.verification-find} covers the find-set operation and various path-compression techniques such as path splitting and path halving.
Section \ref{section.verification-union} is concerned with the union-sets operation, in particular, using the union-by-rank strategy which involves working with natural numbers in a relational framework.
In Section \ref{section.related-work} we discuss related work and limitations.

In addition to the main results, which prove the correctness of various algorithms for disjoint-set forests, throughout the paper we include a number of theorems capturing properties of associative array access, directed forests, reachability in graphs, graph components and Peano structures.
Some of these facts are used in the correctness proofs while others arise from studying the involved concepts.
They all build on and become part of a formally verified library of (currently several thousand) results related to Kleene algebras, relation algebras and similar algebraic structures, which is part of the Archive of Formal Proofs.
These libraries have been developed over many years in previous work and are structured into a hierarchy of algebras; each result is usually proved in the most general structure to avoid unnecessary assumptions and make it more widely applicable.
Such theory-building not only helps to understand concepts but also enables effective automated theorem proving support during a development and in future developments.
It is often not clear in advance which results might turn out to be useful later; automated tools such as Sledgehammer \cite{PaulsonBlanchette2010} can filter the large collection and identify relevant results if and when they are needed.
Hence in this paper we present not just results that are actually used in the overall correctness proof; in fact it would require tool support to gather this information and the result might not give much insight as automatically constructed proofs often differ from manually constructed ones.

The present paper is a revised and extended version of \cite{Guttmann2020a}.
Section 6 of the conference version has been split into current Sections \ref{section.verification-init}--\ref{section.verification-union} to accomodate new material.
The extensions comprise new Sections \ref{section.motivation}, \ref{section.init-sets}, \ref{section.path-splitting}, \ref{section.path-halving}, \ref{section.peano}--\ref{section.union-by-rank} and \ref{section.limitations} and Appendix \ref{section.proofs}, as well as additions to Sections \ref{section.arrays}, \ref{section.path-compression}, \ref{section.find-set-path-compression} and \ref{section.union-sets} and Appendix \ref{section.forest-properties}.
They cover additional properties of read and write operations on arrays, the initialisation of disjoint-set forests, a number of abstractions, the precise effect of path compression, path splitting, path halving, results on relational Peano structures and union-by-rank.
All new results have been formally proved in Isabelle/HOL requiring a substantial extension of the theories developed for \cite{Guttmann2020a}.

\section{Motivation}
\label{section.motivation}

In this section we motivate the relation-algebraic approach to program correctness.
We argue that relation algebras facilitate program development and verification by supporting concise high-level specifications, equational reasoning and generalisation.

\medskip
In the introduction we have summarised how relations and algebras have been used to model programs and their development \cite{BackWright1998,HoareHe1998,Maddux1996,SchmidtStroehlein1989,BackWright1999,Kozen1997,Kozen1994,BackhouseCarre1975,Berghammer1999,BerghammerKargerWolf1998,BerghammerStruth2010,GondranMinoux2008,HoefnerMoeller2012,Moeller1993,BirdMoor1997}.
The underlying idea is to use algebraic equations and inequalities instead of logical formulas.
For example, recall that relation $R$ is transitive if
\begin{align*}
  \forall x, y, z : (x,y) \in R \wedge (y,z) \in R \implies (x,z) \in R
\end{align*}
In relation algebras this is expressed as $R \comp R \subseteq R$ using relational composition.
For another example, directed graph $R$ is acyclic if
\begin{align*}
  \neg \exists n : \exists x_0,\dots,x_n : (x_n,x_0) \in R \wedge \forall i : 0 \leq i < n \implies (x_i,x_{i+1}) \in R
\end{align*}
In relation algebras this is expressed as $R^+ \subseteq \cpl{\unit}$ stating that the transitive closure of $R$ is included in the complement of the identity relation.
These examples demonstrate how eliminating quantifiers can lead to concise specifications.
The resulting algebraic expressions allow equational proofs instead of pointwise arguments.
This is similar to how higher-order functions and pointfree properties raise the level of abstraction in functional programming: for example, the map fusion law can be expressed as $\map \; f \circ \map \; g = \map \; (f \circ g)$ rather than the pointwise $\map \; f \; (\map \; g \; \xs) = \map \; (f \circ g) \; \xs$.

To further motivate the relational approach, we discuss two examples from algorithm development.
First, the maximum segment sum problem computes the maximum sum of a contiguous sublist of a given list of numbers.
A relational specification is given by $\Lambda(\mathit{suffix} \comp \mathit{prefix} \comp \mathit{sum}) \comp \mathit{max}$ \cite{Mu2008}.
Here, $\mathit{prefix}$ and $\mathit{suffix}$ are the prefix and suffix relations on lists, respectively; $\mathit{sum}$ computes the sum of a list of numbers; the power transpose $\Lambda$ converts a relation into a set-valued mapping; $\mathit{max}$ picks the maximum element from a set of numbers.
Thus the specification generates all contiguous sublists (prefixes of suffixes of a list), computes the sum of each, and takes the maximum of the resulting numbers.
This concise specification is easily seen to be valid, but a straight-forward implementation is not optimal.
In \cite{Mu2008}, an optimal linear-time, constant-space algorithm is derived from this specification using equational reasoning.
Correctness of the efficient algorithm follows from the method of construction by semantics-preserving program derivation.
This is based on algebraic properties of the relations involved in the specification.
Thus the specification avoids quantification over list indices and facilitates the application of general optimisation strategies, and the correctness proof avoids low-level reasoning about list indices.

Second, the minimum spanning tree problem computes a tree with minimum total weight that spans a given weighted graph, where to span requires the tree to be an inclusion-maximal acyclic subgraph.
This problem and a number of efficient algorithms for it are well known \cite{CormenLeisersonRivest1990}.
A relation-algebraic specification and implementation have been given in \cite{Guttmann2018b}, and correctness of the implementation has been formally verified in Isabelle/HOL.
Because correctness is entirely proved in extensions of relation algebras, different instances of these algebras can be constructed to solve different problems.
Hence the same correctness proof covers the minimum-weight spanning tree problem, the minimum bottleneck spanning tree problem and further optimisation problems; see \cite{Guttmann2018b} for details.
This is similar to how Warshall's transitive closure algorithm and Floyd's all-pairs shortest paths algorithm have been generalised to semirings, which cover many other instances.
In both cases, the minimum spanning tree and the shortest path, the generalisation relies on using algebras to model the problem, implement the solution and perform its correctness proof.

The aim of this paper is to show that the relation-algebraic approach to program correctness can be extended by associative arrays, thereby providing access to efficient implementations when needed while maintaining the above benefits of high-level specification, equational reasoning and generalisation due to algebras.

\section{Relation algebras and Kleene algebras}
\label{section.relations}

This section presents the algebraic structures used in this development including relation algebras and Kleene algebras and basic properties \cite{Kozen1994,SchmidtStroehlein1989,Tarski1941}.

A \emph{semilattice} $(S,\join)$ is a set $S$ with a binary operation $\join$ that is associative, commutative and idempotent.
In a semilattice the binary relation $\lleq$ defined by $x \lleq y \iff x \join y = y$ is a partial order called the \emph{semilattice order}.
The operation $\join$ is $\lleq$-isotone and gives the $\lleq$-least upper bound or join of two elements.

A \emph{bounded semilattice} $(S,\join,\lbot)$ is a semilattice $(S,\join)$ with a constant $\lbot$ that is a unit of $\join$.
It follows that $\lbot$ is the $\lleq$-least element of $S$.

A \emph{lattice} $(S,\join,\meet)$ comprises two semilattices $(S,\join)$ and $(S,\meet)$ such that the absorption laws $x \join (x \meet y) = x = x \meet (x \join y)$ hold.
The operation $\meet$ is $\lleq$-isotone and gives the $\lleq$-greatest lower bound or meet of two elements.

A \emph{bounded lattice} $(S,\join,\meet,\lbot,\ltop)$ comprises two bounded semilattices $(S,\join,\lbot)$ and $(S,\meet,\ltop)$ such that $(S,\join,\meet)$ is a lattice.
It follows that $\ltop$ is the $\lleq$-greatest element of $S$ and a zero of $\join$, and that $\lbot$ is a zero of $\meet$.

A lattice is \emph{distributive} if the law $x \join (y \meet z) = (x \join y) \meet (x \join z)$ holds.
In a lattice this law is equivalent to its dual $x \meet (y \join z) = (x \meet y) \join (x \meet z)$.

A \emph{Boolean algebra} $(S,\join,\meet,\cplop,\lbot,\ltop)$ is a bounded lattice $(S,\join,\meet,\lbot,\ltop)$ that is distributive with a unary operation $\cplop$ satisfying the laws $x \join \cpl{x} = \ltop$ and $x \meet \cpl{x} = \lbot$.
The operation $\cplop$ is $\lleq$-antitone.
We write $x - y$ for $x \meet \cpl{y}$.

A \emph{monoid} $(S,\comp,\unit)$ is a set $S$ with a binary composition operation $\comp$ that is associative and a constant $\unit$ that is a left unit and a right unit of $\comp$.

An \emph{idempotent semiring} $(S,\join,\comp,\lbot,\unit)$ is a bounded semilattice $(S,\join,\lbot)$ and a monoid $(S,\comp,\unit)$ such that $\comp$ distributes over $\join$ and $\lbot$ is a left zero and a right zero of $\comp$.
The operation $\comp$ is $\lleq$-isotone.

\medskip
A \emph{relation algebra} $(S,\join,\meet,\comp,\cplop,\cnvop,\lbot,\!\ltop,\unit)$ is a Boolean algebra $(S,\join,\meet,\cplop,\lbot,\!\ltop)$ and an idempotent semiring $(S,\join,\comp,\lbot,\unit)$ with a unary transposition operation $\cnvop$ satisfying the laws:
\begin{align*}
  \cnv{(x \join y)} & = \cnv{x} \join \cnv{y} & \cnv{\cnv{x}} & = x \\
  \cnv{(x \comp y)} & = \cnv{y} \comp \cnv{x} & (x \comp y) \meet z & \lleq x \comp (y \meet (\cnv{x} \comp z))
\end{align*}
It follows that the operation $\cnvop$ is $\lleq$-isotone.
A relation algebra satisfies the \emph{Tarski rule} if $\ltop \comp x \comp \ltop = \ltop$ for each $x \neq \lbot$.

\medskip
A \emph{Kleene algebra} $(S,\join,\comp,\astop,\lbot,\unit)$ is an idempotent semiring $(S,\join,\comp,\lbot,\unit)$ with a unary iteration operation $\astop$ satisfying the laws:
\begin{align*}
  \unit \join (y \comp y^*) & = y^* & z \join (y \comp x) \lleq x & \implies y^* \comp z \lleq x \\
  \unit \join (y^* \comp y) & = y^* & z \join (x \comp y) \lleq x & \implies z \comp y^* \lleq x
\end{align*}
The operation $\astop$ is a closure operation: it is idempotent, $\lleq$-isotone and $\lleq$-increasing, that is, $x \lleq x^*$.
It describes finite iterations with zero or more steps; the related operation $x^+ = x \comp x^*$ describes finite iterations with one or more steps.
In particular, $y^* \comp z$ and $z \comp y^*$ are the $\lleq$-least fixpoints of the functions $\lambda x . (y \comp x) \join z$ and $\lambda x . (x \comp y) \join z$, respectively.
Moreover $(x \join y)^* = x^* \comp (y \comp x^*)^*$ and $x \comp (y \comp x)^* = (x \comp y)^* \comp x$.

\medskip
A \emph{Kleene relation algebra} $(S,\join,\meet,\comp,\cplop,\cnvop,\astop,\lbot,\ltop,\unit)$ comprises a Kleene algebra $(S,\join,\comp,\astop,\lbot,\unit)$ and a relation algebra $(S,\join,\meet,\comp,\cplop,\cnvop,\lbot,\ltop,\unit)$.
It follows that $\cnv{x^*} = \cnv{x}\astop$.

An element $x \in S$ of a relation algebra $S$ is called
  \emph{reflexive} if $\unit \lleq x$,
  \emph{transitive} if $x \comp x \lleq x$,
  \emph{symmetric} if $\cnv{x} = x$,
  an \emph{equivalence} if $x$ is reflexive and transitive and symmetric,
  \emph{total} if $\unit \lleq x \comp \cnv{x}$,
  \emph{surjective} if $\unit \lleq \cnv{x} \comp x$,
  \emph{univalent} if $\cnv{x} \comp x \lleq \unit$,
  \emph{injective} if $x \comp \cnv{x} \lleq \unit$,
  \emph{bijective} if $x$ is injective and surjective,
  a \emph{mapping} if $x$ is univalent and total,
  a \emph{vector} if $x \comp \ltop = x$,
  a \emph{point} if $x$ is a vector and bijective,
  and an \emph{arc} if $x \comp \ltop$ and $\cnv{x} \comp \ltop$ are bijective.

An element $x \in S$ of a Kleene relation algebra $S$ is called
  \emph{acyclic} if $x^+ \lleq \cpl{\unit}$.

In this paper we work in a Kleene relation algebra $S$ that satisfies the Tarski rule.
For proving termination of programs we assume that $S$ is finite.
In some cases it would suffice to require that $S$ is complete at the cost of strengthening acyclic to progressively finite \cite{SchmidtStroehlein1989}; we do not pursue this because the algorithms are usually applied to finite disjoint-set forests.

The main model of Kleene relation algebras are binary relations over a set $A$, that is, subsets of $A \times A$.
In this model $\join$ is union, $\meet$ is intersection, $\cplop$ is complement, $\lleq$ is subset, $\lbot$ is the empty set, $\ltop$ is $A \times A$, $\comp$ is relational composition, $\cnvop$ is relational transposition, $\unit$ is the identity relation, $\astop$ is reflexive transitive closure, $\plusop$ is transitive closure, and the Tarski rule holds.

We finally characterise vectors, points and arcs among the binary relations over $A$.
A vector is a relation $B \times A$ for a subset $B \subseteq A$; hence vectors represent subsets of the base set such as a set of nodes in a graph.
Vectors correspond to row-constant matrices and are also called `conditions' in the literature.
A point is a relation $\{a\} \times A$ for an element $a \in A$; hence points represent elements of the base set such as nodes in a graph.
In particular, a point is injective, surjective and a vector, but not univalent (if $A$ has more than two elements).
An arc is a relation $\{(a,b)\}$ for elements $a, b \in A$; hence arcs represent pairs of elements from the base set such as edges in a graph.

\section{Associative array access}
\label{section.arrays}

An array maps indices to values and therefore can be modelled as a binary relation between two sets.
Under our assumption that indices and values come from the same set $A$, we can use binary relations on $A$ and work with these using relation algebra.
Because an array associates exactly one value to every index, the relation is a mapping in the relation-algebraic sense, that is, univalent and total.
A relation that is just univalent corresponds to a partially defined array which associates at most one value to every index.
We use such relations to model arrays which have not yet been fully initialised.
An index or a value is an element of $A$, which can be modelled in relation algebras as a point.
A relation that is just a vector corresponds to a set of indices or values.

\medskip
These observations underlie the following simple semantics of array access.
Let $x$, $y$ and $z$ be elements of a relation algebra such that $y$ and $z$ are points.
The element $x$ models the associative array, $y$ corresponds to an index and $z$ corresponds to a value.
The array $\awrite{x}{y}{z}$ obtained by updating array $x$ at index $y$ to new value $z$ is:
\begin{align*}
  \awrite{x}{y}{z} = (y \meet \cnv{z}) \join (\cpl{y} \meet x)
\end{align*}
To understand this definition it is helpful to consider the matrix representation of the relation modelling the array $x$.
A vector describes a set of rows of the matrix and a point describes a single row.
The point $y$ refers to the row at the corresponding index.
Its complement $\cpl{y}$ refers to all the other rows.
The formula $\cpl{y} \meet x$ specifies that in all other rows $x$ is left unchanged.
The formula $y \meet \cnv{z}$ specifies that row $y$ is updated to value $z$.
Since $z$ is a point, which refers to a row, we take its transposition $\cnv{z}$, which refers to the column of the matrix at the corresponding value.
In terms of binary relations, $y \meet \cnv{z}$ constructs a relation containing a single pair of the index $y$ and the value $z$.
In relation algebras $y \meet \cnv{z} = y \comp \cnv{z}$ is an arc for points $y$ and $z$.

\medskip
For example, consider the following relations $x$, $y$ and $z$ on $A = \{ 1, 2, 3 \}$ given as Boolean matrices:
\begin{align*}
  x & = \matt{0}{0}{1}{0}{1}{0}{0}{0}{0} & y & = \matt{0}{0}{0}{1}{1}{1}{0}{0}{0} & z & = \matt{1}{1}{1}{0}{0}{0}{0}{0}{0}
\end{align*}
Relation $x$ represents a partially defined array that maps index $1$ to value $3$ and index $2$ to value $2$, point $y$ represents index $2$ and point $z$ represents value $1$.
The updated array still maps index $1$ to value $3$, but maps index $2$ to value $1$:
\begin{align*}
  y \meet \cnv{z} & = \matt{0}{0}{0}{1}{0}{0}{0}{0}{0} & \cpl{y} \meet x & = \matt{0}{0}{1}{0}{0}{0}{0}{0}{0} & \awrite{x}{y}{z} & = \matt{0}{0}{1}{1}{0}{0}{0}{0}{0}
\end{align*}

Reading the value $\aread{x}{y}$ of the associative array $x$ at index $y$ is done by:
\begin{align*}
  \aread{x}{y} = \cnv{x} \comp y
\end{align*}
The composition of a relation with a vector always gives a vector.
If $x$ is interpreted as a transition relation, $\cnv{x} \comp y$ is a vector corresponding to the successors of the point $y$ under a transition step according to $x$.
In the matrix representation of an array, this is just the value of $x$ at row $y$.
If the array associates exactly one value to every index, the result is the unique value associated with index $y$, represented as a point.

\medskip
Continuing the previous example, the value of $x$ at index $y$ is $2$ and the value of $x$ at index $z$ is $3$:
\begin{align*}
  \cnv{x} & = \matt{0}{0}{0}{0}{1}{0}{1}{0}{0} & \aread{x}{y} & = \matt{0}{0}{0}{1}{1}{1}{0}{0}{0} & \aread{x}{z} & = \matt{0}{0}{0}{0}{0}{0}{1}{1}{1}
\end{align*}

In the remainder of this paper, we use the above equational definitions of array write and read without implicitly assuming that $y$ and $z$ are points.
This allows updating the values of several indices at once and other general ways of writing and reading which can be useful for specification and reasoning.

\medskip
The following result shows basic preservation properties of write and read operations on arrays.
Part 1 is similar to \cite[Lemma 2.7]{Moeller1993b}.

\begin{thm}
  \label{theorem.awrite-aread}
  \begin{enumerate}
  \item[]
  \item $\awrite{x}{y}{z}$ is univalent if $x$ is univalent, $y$ is a vector and $z$ is injective.\afplemma{update_univalent}
  \item $\awrite{x}{y}{z}$ is total if $x$ is total, $y$ is a vector and $z$ is surjective.\afplemma{update_total}
  \item $\awrite{x}{y}{z}$ is a mapping if $x$ is a mapping, $y$ is a vector and $z$ is bijective.\afplemma{update_mapping}
        \label{theorem.awrite-aread.update-mapping}
  \item $\aread{x}{y}$ is injective if $x$ is univalent and $y$ is injective.\afplemma{read_injective}
  \item $\aread{x}{y}$ is surjective if $x$ is total and $y$ is surjective.\afplemma{read_surjective}
  \item $\aread{x}{y}$ is bijective if $x$ is a mapping and $y$ is bijective.\afplemma{read_bijective}
  \item $\aread{x}{y}$ is a point if $x$ is a mapping and $y$ is a point.\afplemma{read_point}
        \label{theorem.awrite-aread.aread-point}
  \item $\aread{x}{y} = z \iff y \meet x = y \comp \cnv{z}$ if $y$ and $z$ are points.\afplemma{update_postcondition}
        \label{theorem.awrite-aread.aread-iff}
  \end{enumerate}
\end{thm}

\begin{prf}
  See Appendix \ref{section.proofs} for a proof of Theorem \ref{theorem.awrite-aread}.\ref{theorem.awrite-aread.aread-iff} and \cite{Guttmann2020c} for all proofs.
\end{prf}

The following result shows further properties of write and read operations on arrays, which are helpful for verifying the correctness of operations on disjoint-set forests.
Parts 1--5 are the put-get, put-put and get-put lens properties \cite{FosterGreenwaldMoorePierceSchmitt2007} and the array independence requirements of \cite{BackWright1998}.
Parts 6--7 combine and split updates with the same value.
Parts 8--10 swap the values $\aread{x}{y}$ and $\aread{x}{z}$.

\begin{thm}
  \label{theorem.awrite-aread-2}
  \begin{enumerate}
  \item[]
  \item $\aread{\awrite{x}{y}{z}}{y} = z$ if $y$ is a surjective vector and $z$ is a vector.\afplemma{put_get}
  \item $\aread{\awrite{x}{y}{z}}{u} = \aread{x}{u}$ if $u \lleq \cpl{y}$ and $y$ is a vector.\afplemma{put_get_different_vector}
  \item $\awrite{\awrite{x}{y}{z}}{y}{u} = \awrite{x}{y}{u}$.\afplemma{put_put}
  \item $\awrite{\awrite{x}{y}{z}}{u}{v} = \awrite{\awrite{x}{u}{v}}{y}{z}$ if $u$ and $y$ are vectors with $u \meet y = \lbot$.\afplemma{put_put_different_vector}
  \item $\awrite{x}{y}{\aread{x}{y}} = x$ if $y$ is a point.\afplemma{get_put}
  \item $\awrite{\awrite{x}{y}{z}}{u}{z} = \awrite{x}{y \join u}{z}$.\afplemma{update_same}
  \item $\awrite{x}{y}{z} = \awrite{\awrite{x}{y - w}{z}}{y \meet w}{z}$.\afplemma{update_split}
  \item $\awrite{\awrite{x}{y}{\aread{x}{z}}}{z}{\aread{x}{y}}$ is injective if $x$ is injective, $y$ is a point and $z$ is an injective vector.\afplemma{update_injective_swap}
        \label{theorem.awrite-aread-2.update-injective-swap}
  \item $\awrite{\awrite{x}{y}{\aread{x}{z}}}{z}{\aread{x}{y}}$ is univalent if $x$ is univalent and $y$ and $z$ are injective vectors.\afplemma{update_univalent_swap}
  \item $\awrite{\awrite{x}{y}{\aread{x}{z}}}{z}{\aread{x}{y}}$ is a mapping if $x$ is a mapping and $y$ and $z$ are points.\afplemma{update_mapping_swap}
  \item $\awrite{x}{\lbot}{z} = x$.\afplemma{update_bot}
  \item $\awrite{x}{\ltop}{z} = \cnv{z}$.\afplemma{update_top}
  \item $\awrite{x}{y}{z} \lleq x \join \cnv{z}$.\afplemma{update_ub}
        \label{theorem.awrite-aread-2.update-ub}
  \item $\awrite{x}{y}{\cnv{z}} = x$ if $y \lleq u$ and $u \meet x = u \meet z$ for some $u$.\afplemma{update_same_sub}
  \item $\awrite{x}{y}{\aread{z}{y}} = \awrite{x}{y}{\cnv{z}}$ if $y$ is a point.\afplemma{update_point_get}
  \item $x \meet y$ is an arc if $x$ is a mapping and $y$ is a point.\afplemma{mapping_inf_point_arc}
  \end{enumerate}
\end{thm}

\begin{prf}
  See Appendix \ref{section.proofs} for a proof of Theorem \ref{theorem.awrite-aread-2}.\ref{theorem.awrite-aread-2.update-injective-swap} and \cite{Guttmann2020c} for all proofs.
\end{prf}

We conclude this section by discussing the representation of array indices using points.
Restrictions with vectors, such as $y \meet \cnv{z}$ and $\cpl{y} \meet x$ in an array update, can also be obtained by composing with partial identities, as in $(y \meet \unit) \comp \cnv{z}$ and $(\cpl{y} \meet \unit) \comp x$.
In relation algebras, partial identities and vectors provide equivalent ways of representing sets which are readily converted to each other.
Yet another way to represent sets is to use the transposition of a vector.
Each of these representations has advantages in some situations and disadvantages in others.
For example, using the transposition of a point for array indices would simplify array read $\aread{x}{y}$ to a simple composition, but would complicate array update; representing sets by partial identities would make restrictions from the left and from the right more symmetric and would allow some results such as Theorem \ref{theorem.path-compression}.\ref{theorem.path-compression.optimise-1} in Appendix \ref{section.forest-properties} to be generalised to Kleene algebras, but would require an extra domain operation to identify certain sets of nodes in a graph.
It is not the purpose of this paper to study advantages and disadvantages of the various representations; any of them would be fine as regards the contributions of this paper.

\section{Disjoint sets}
\label{section.disjoint-sets}

A disjoint-set data structure keeps track of a set of elements that is partitioned into disjoint sets \cite{GallerFisher1964}.
The basic operations are to initialise elements to be in their own singleton sets, to form the union of two sets and to look up which set an element belongs to.

The semantics of a disjoint-set data structure with elements from $A$ is an equivalence relation on $A$.
The disjoint sets are just the equivalence classes of the relation.
A particular representative from each class identifies a set.

An element of a relation algebra is an equivalence if it is reflexive, transitive and symmetric.
The $\lleq$-least equivalence is the identity relation $\unit$.
The $\lleq$-greatest equivalence is the universal relation $\ltop$.
Equivalences are closed under the operations $\meet$ and $\cnvop$ and, in Kleene relation algebras, under $\astop$ and $\plusop$.

\medskip
The \emph{equivalence closure} of a relation $x$ is the smallest equivalence relation containing $x$.
According to \cite{SchmidtStroehlein1989} it is given by:
\begin{align*}
  \wcc(x) = (x \join \cnv{x})^*
\end{align*}
Interpreting the relation $x$ as a directed graph, the equivalence closure represents the weakly-connected components of $x$ (hence the short name $\wcc$), which are obtained by reachability while ignoring the direction of edges.
Properties of $\wcc$ are given in the following result.
Part 3 shows a Galois connection between $\wcc$ and the forgetful functor mapping equivalences to relations, which is a consequence of being a closure operation.

\begin{thm}
  \label{theorem.wcc}
  \begin{enumerate}
  \item[]
  \item $\wcc(x)$ is an equivalence.\afplemma{wcc_equivalence}
  \item $\wcc$ is a closure operation.\afplemma{wcc_increasing}
        \label{theorem.wcc.closure}
  \item $\wcc(x) \lleq \wcc(y)$ if and only if $x \lleq \wcc(y)$.\afplemma{wcc_below_wcc}
        \label{theorem.wcc.galois}
  \item $\wcc(\lbot) = \wcc(\unit) = \unit$.\afplemma{wcc_bot}
  \item $\wcc(\ltop) = \ltop$.\afplemma{wcc_top}
  \item $\wcc(x \join \unit) = \wcc(x - \unit) = \wcc(x)$.\afplemma{wcc_with_loops}
  \item $\wcc(x \join y) = \wcc(x \join \wcc(y))$.\afplemma{wcc_sup_wcc}
        \label{theorem.wcc.join-wcc}
  \end{enumerate}
\end{thm}

\begin{prf}
  See Appendix \ref{section.proofs} for a proof of Theorem \ref{theorem.wcc}.\ref{theorem.wcc.closure} and \cite{Guttmann2020c} for all proofs.
\end{prf}

Following \cite{CormenLeisersonRivest1990} we implement the data structure as a disjoint-set forest.
These are directed forests in the graph-theoretical sense except the root of each tree has a self-loop, that is, an edge to itself.
Each equivalence class corresponds to a tree in the forest.
Singleton sets correspond to singleton trees, which contain one node with a self-loop.
Each tree in the forest has a root and is directed.
Each node in a tree has a unique parent node; the root is its own parent.
The root of a tree represents the corresponding equivalence class.
An edge from a node to its parent points towards the root of the tree, which can be reached by successively following parents.

Disjoint-set forests can be modelled in Kleene relation algebras as follows.
An element $x \in S$ of a Kleene relation algebra $S$ is called a \emph{forest} if $x$ is a mapping and $x - \unit$ is acyclic.
Requiring $x$ to be a mapping ensures that each node has a unique parent.
It remains to ensure that there are no cycles.
We cannot require that $x$ is acyclic because every root has itself as its parent, which corresponds to a self-loop in the graph.
However, $x - \unit$ removes all self-loops, so we require that the result is acyclic.
Related helpful lemmas are $x^*= (x - \unit)^*$ and $x^* - \unit = x^+ - \unit$.

\medskip
In a forest $x$, it is possible to reach from a node every node in the same component tree by going towards its root and then back to the desired node.
This defines a relation $\fc(x)$ on the nodes of the forest, namely the relation of being in the same component ($\fc$ is short for forest-components):
\begin{align*}
  \fc(x) = x^* \comp \cnv{x}\astop
\end{align*}
Properties of this construction are given in the following result.
Part \ref{theorem.component.wcc-fc} shows that $\fc$ and $\wcc$ coincide for forests.

\begin{thm}
  \label{theorem.component}
  \begin{enumerate}
  \item[]
  \item $\fc(x)$ is an equivalence if $x$ is univalent.\afplemma{fc_equivalence}
        \label{theorem.component.equivalence}
  \item $\fc$ is $\lleq$-increasing.\afplemma{fc_increasing}
  \item $\fc$ is $\lleq$-isotone.\afplemma{fc_isotone}
  \item $\fc(\fc(x)) = \fc(x)$ if $x$ is univalent.\afplemma{fc_idempotent}
  \item $\fc(x)^* = \fc(x)^+ = \fc(x)$ if $x$ is univalent.\afplemma{fc_star}
  \item $\fc(\lbot) = \fc(\unit) = \unit$.\afplemma{fc_bot}
  \item $\fc(\ltop) = \ltop$.\afplemma{fc_top}
  \item $\fc(x) = \wcc(x)$ if $x$ is univalent.\afplemma{fc_wcc}
        \label{theorem.component.wcc-fc}
  \end{enumerate}
\end{thm}

\begin{prf}
  See Appendix \ref{section.proofs} for a proof of Theorem \ref{theorem.component}.\ref{theorem.component.equivalence} and \cite{Guttmann2020c} for all proofs.
\end{prf}

Functions $\wcc$ and $\fc$ can be used to provide abstract specifications for operations on disjoint-set forests.
The abstraction function $\fc$ maps a forest to the equivalence relation it represents.
Operations such as path compression change the forest but not the represented equivalence relation, which is expressed by keeping the value under $\fc$ constant.
Operations such as union-sets change the equivalence relation, which is expressed by computing the new equivalence using $\wcc$.
The specifications will be given implicitly by the pre- and postconditions of the verified implementations in the following sections.
Alternative methods to construct formally correct programs start with a specification and proceed by program transformations and data refinement; for example, see \cite{DerrickBoiten2018,Jones1980,Partsch1990}.
This paper follows the program verification method, which proves correctness of a given implementation with respect to a given specification.

\section{Verifying disjoint-set forests -- initialisation}
\label{section.verification-init}

For implementing the operations on disjoint-set forests and verifying their correctness we use a Hoare-logic library of Isabelle/HOL \cite{Nipkow1998,Nipkow2002}, which we have extended from partial correctness to total correctness \cite{Guttmann2018c}.
The library supports while-programs, which have to be annotated with a precondition, a postcondition, and an invariant and a variant for each while-loop.
From this, verification conditions are automatically generated.
Soundness of the Hoare-logic rules is verified with respect to a standard operational semantics of the while-programs.

Program variables can range over arbitrary HOL types.
We write programs in the context of an Isabelle/HOL class specifying the axioms of Kleene relation algebras, the Tarski rule and a finite universe for total correctness.
Hence program variables range over elements from the universe of the class, which models the corresponding algebraic structure.
Reasoning about these variables to discharge verification conditions is performed in the same context using existing libraries for Kleene algebras and relation algebras and newly derived theorems.
The values of the program variables can be interpreted as binary relations and represented by Boolean matrices, like the examples in Section \ref{section.arrays}, but other interpretations are possible as long as they satisfy the given axioms.
The Boolean matrix model has been formalised in Isabelle/HOL and proved to satisfy the axioms to ensure their soundness.

While-programs supported by the Hoare-logic library feature while-loops, conditionals, sequential composition and assignments as basic statements.
We introduce new notation for array read and write operations, which are automatically translated to basic relation-algebraic expressions according to Section \ref{section.arrays}.
The assignment $\aset{x}{y}{z}$ is translated to the assignment $x := \awrite{x}{y}{z}$.
The read expression $\aread{x}{y}$ can be used directly on the right-hand side of assignments and in conditions, except we modify its syntax to $\aread{x}{[y]}$ to avoid ambiguity with list syntax in Isabelle/HOL.
This paper uses $\aread{x}{y}$ to improve readability.
In the Boolean matrix model of relations, arrays representing forests are in one-to-one correspondence with finite maps.

This section covers verification of the initialisation phase of disjoint-set forests.
The find and union operations are covered in Sections \ref{section.verification-find} and \ref{section.verification-union}, respectively.\vspace*{-1mm}

\subsection{The make-set operation}
\label{section.make-set}

As a warm-up we implement the make-set operation of disjoint-set forests and prove its correctness.
It is usually applied to each element when the data structure is initialised.
Until the initialisation is complete, the underlying associative array is partial.
Make-set puts an element $x$ into its own singleton equivalence class by setting the parent of $x$ to itself, which creates a singleton tree:
\begin{numquote}\small
  \textbf{theorem} make\_set:\afptheorem{make_set} \\
  \ind ``VARS $p$ \\
  \ind \ind [ point $x$ $\wedge$ $p_0$ = $p$ ] \\
  \ind \ind $\aset{p}{x}{x}$ \\
  \ind \ind [ make\_set\_postcondition $p$ $x$ $p_0$ ]'' \\
  \ind \textbf{apply} vcg\_tc\_simp \\
  \ind \textbf{by} (simp add: ...) -- names of four lemmas omitted
\end{numquote}
Lines 2--5 combine an implementation (line 4) with a specification comprising a frame (line 2), a precondition (line 3) and a postcondition (line 5) as in \cite{Morgan1994}.
The specification can also be understood as a Hoare triple extended to include the frame whose scope is the entire triple.
The theorem states that the implementation is totally correct with respect to the specification.

\medskip
Line 2 declares variables that are changed by the program and therefore need to be part of the state, in this case only $p$ which contains the parent array.
The variables $x$ and $p_0$ are universally quantified variables of the theorem; because they are not changed they do not need to be part of the state.
The variable $p_0$ transports the initial value of $p$ to the postcondition, where it can be related to the final value of $p$.
Line 3 gives the precondition, which requires $x$ to be a point, representing an element of the set partitioned by the data structure.
Line 4 updates the parent array to make $x$ the root of a tree.
Line 5 gives the postcondition, which is discussed below.
Line 6 generates the verification condition, which for this small program is a single goal, and applies some simplifications to it:
\begin{quote}
  point $x$ $\wedge$ $p_0$ = $p$ $\implies$ make\_set\_postcondition $\awrite{p}{x}{x}$ $x$ $p$
\end{quote}
Line 7 proves this goal by invoking the simplifier with additional lemmas.
The postcondition has two parts:\vspace*{-1mm}
\begin{quote}
  make\_set\_postcondition $p$ $x$ $p_0$ $\iff$ $x \meet p = x \comp \cnv{x} \wedge \cpl{x} \meet p = \cpl{x} \meet p_0$
\end{quote}
The first condition $x \meet p = x \comp \cnv{x}$ states that the parent array contains $x$ at index $x$.
It is equivalent to $\aread{p}{x} = x$ by Theorem \ref{theorem.awrite-aread}.\ref{theorem.awrite-aread.aread-iff}.
The second condition $\cpl{x} \meet p = \cpl{x} \meet p_0$ states that the parent array remains unchanged at all indices different from $x$.
Both conditions together are equivalent to $p = \awrite{p_0}{x}{x}$.

The precondition and postcondition can be strengthened by adding $p \lleq \unit$.
As a consequence, when a disjoint-set forest is initialised each equivalence class constructed by make-set is a singleton.

The method vcg\_tc\_simp generates conditions that prove total correctness.
Since the above program does not contain any while-loops, there are no conditions related to its termination.

\medskip
We use a basic Hoare-logic library which does not support the definition of procedures.
So that other programs can use a disjoint-set operation such as make-set, we extract an Isabelle/HOL function from the above proof using a technique of \cite{Guttmann2018c}.
Specifically, the above total-correctness theorem implies:
\begin{flushleft}
  \textbf{lemma} make\_set\_exists: ``point $x$ $\implies$ $\exists p'$ . make\_set\_postcondition $p'$ $x$ $p$'' \\
  \ind \textbf{using} tc\_extract\_function make\_set \textbf{by} blast
\end{flushleft}
This is a consequence of how total correctness is defined on the underlying operational semantics.
Hence we can introduce the following Isabelle/HOL function:
\begin{flushleft}
  \textbf{definition} ``make\_set $p$ $x$ = (SOME $p'$ . make\_set\_postcondition $p'$ $x$ $p$)''
\end{flushleft}
The construct SOME $y$ .\ $P(y)$ yields some element $y$ that satisfies $P(y)$.
In order to reason about this function in other programs we derive the following property:
\begin{flushleft}
  \textbf{lemma} make\_set\_function: \\
  \ind \textbf{assumes} ``point $x$'' \textbf{and} ``$p'$ = make\_set $p$ $x$'' \\
  \ind \textbf{shows} ``make\_set\_postcondition $p'$ $x$ $p$'' \\
  \ind -- proof omitted
\end{flushleft}
Ideally, the above technique or another construction that provides an Isabelle/HOL function and a corresponding pre-/postcondition result should be part of the Hoare-logic library.

\subsection{Initialising disjoint-set forests}
\label{section.init-sets}

A disjoint-set forest is initialised by applying the make-set operation to each element.
In \cite{CormenLeisersonRivest1990} this is achieved by a simple for-each-loop.
Since the Hoare-logic library only supports while-loops we first introduce a suitable abstraction.

\medskip
Iteration through the elements is implemented using a working set of elements that have not been considered yet from which an element is selected and removed in each step.
Sets of elements are represented as vectors.
The selected element is represented as a point and obtained by the operation choose\_point with the following axioms \cite{Berghammer1999}:
\begin{enumerate}
\itemsep=0.8pt
\item vector $x$ $\wedge$ $x \neq \lbot$ $\implies$ point (choose\_point $x$)
\item choose\_point $x$ $\lleq$ $x$
\end{enumerate}
The first axiom specifies that the result of choose\_point $x$ is one element if $x$ is a non-empty set of elements.
The second axiom specifies that the resulting element is contained in $x$.
Using this operation we define the control structure
\begin{quote}
  FOREACH $x$ \\
  \ind USING $h$ \\
  \ind INV \{ $i$ \} \\
  \ind DO $c$ OD
\end{quote}
as syntactic sugar for
\begin{numquote}
  $h := \ltop$; \\
  WHILE $h \neq \lbot$ \\
  \ind INV \{ vector $h$ $\wedge$ $i$ \} \\
  \ind VAR \{ \down{h} \} \\
  \ind DO \\
  \ind \ind $x$ := choose\_point $h$; \\
  \ind \ind $c$; \\
  \ind \ind $\aset{h}{x}{\lbot}$ \\
  \ind OD
\end{numquote}
Vector $h$ describes the set of all elements that are yet to be processed, which contains all elements at the start according to line 1.
The for-each-loop abstraction makes this vector explicit so the invariant can refer to it.
Point $x$ is the element selected in each iteration in line 6.
In line 8, vector $h$ is interpreted as an array whose value at index $x$ is set to $\lbot$ signifying the removal of $x$ from $h$.
As per line 2, the while-loop terminates when all elements have been processed.

Every while-loop in a program needs to be annotated with an invariant and a variant.
Line 3 augments the invariant of the for-each-loop by the fact that $h$ is a vector.
We also show that the while-loop terminates using the variant in line 4.
The variant is an expression that yields a natural number depending on the program variables.
For the above program, the variant is \down{h}, the number of elements in the algebra below $h$.
The value of this expression decreases after execution of the body of the loop.
Because it is a natural number, it will decrease only a finite number of times which ensures termination of the while-loop.
This works because the algebra is finite.

\medskip
Using a for-each-loop the overall initialisation of a disjoint set forest works as follows:
\begin{numquote}
  \textbf{theorem} init\_sets:\afptheorem{init_sets} \\
  \ind ``VARS $h$ $p$ $x$ \\
  \ind \ind [ True ] \\
  \ind \ind FOREACH $x$ \\
  \ind \ind \ind USING $h$ \\
  \ind \ind \ind INV \{ $p - h = \unit - h$ \} \\
  \ind \ind \ind DO \\
  \ind \ind \ind \ind $p$ := make\_set $p$ $x$ \\
  \ind \ind \ind OD \\
  \ind \ind [ $p = \unit$ $\wedge$ forest $p$ $\wedge$ $h = \lbot$ ]'' \\
  \ind \textbf{apply} vcg\_tc\_simp \\
  \ind -- proof of one verification condition omitted
\end{numquote}
The loop iterates over each element $x$ according to line 4.
Line 8 initialises the disjoint-set forest at that element using the make\_set operation extracted in Section \ref{section.make-set}.

Line 5 declares $h$ to refer to the set of elements yet to be initialised.
The loop invariant in line 6 maintains that all elements in $\cpl{h}$ have already been initialised, that is, the parent of such an element is itself.
According to the postcondition in line 10, after the initialisation the parent array is the identity relation, that is, all elements are roots in the disjoint-set forest.
Line 3 shows that this requires no precondition.

Because the above program contains one loop, three verification conditions are generated: one to establish the loop invariant before execution of the loop, one to maintain the loop invariant across execution of the body of the loop, and one to show the postcondition at the end of the loop.
The built-in simplifier automatically discharges the first and the last of these conditions for init\_sets, and it only remains to show that the invariant is maintained.

An Isabelle/HOL function could be extracted from the correctness of init\_sets as we have done for make\_set in Section \ref{section.make-set}.
We omit this as the effect of init\_sets can be achieved simply by setting $p := \unit$.
The difference is that this assignment initialises the entire array $p$ at once whereas the loop in init\_sets is closer to how initialisation would be implemented in lower-level algorithms.
For example, the algorithm in \cite{CormenLeisersonRivest1990} calls make-set for each node in the graph without specifying an order, just like the for-each-loop abstraction does not specify the order of iteration.

\section{Verifying disjoint-set forests -- find and path-compression}
\label{section.verification-find}

In this section we implement the find-set operation of disjoint-set forests and various path-compression techniques and verify correctness of these algorithms.
Section \ref{section.find-set} covers the basic find-set operation without modifying parent links.
A separate path-compression phase is the topic of Section \ref{section.path-compression}, which also describes the precise effect of this phase.
Both phases are combined and their precise effect is studied in Section \ref{section.find-set-path-compression}.
Two alternative path-compression techniques, namely path splitting and path halving, are discussed in Sections \ref{section.path-splitting} and \ref{section.path-halving}, respectively.

\subsection{The find-set operation}
\label{section.find-set}

We start by implementing the find-set operation of disjoint-set forests and verifying its correctness.
The find-set operation computes the representative of the equivalence class an element belongs to.
We first demonstrate a basic implementation of find-set and then extend it by path compression.
The pseudo-code in \cite{CormenLeisersonRivest1990} uses recursion whereas we use a while-loop.
The find-set operation follows the chain of parents from a node $x$ to the root of its tree:
\begin{numquote}
  \textbf{theorem} find\_set:\afptheorem{find_set} \\
  \ind ``VARS $y$ \\
  \ind \ind [ find\_set\_precondition $p$ $x$ ] \\
  \ind \ind $y := x$; \\
  \ind \ind WHILE $y \neq \aget{p}{y}$ \\
  \ind \ind \ind INV \{ find\_set\_invariant $p$ $x$ $y$ \} \\
  \ind \ind \ind VAR \{ \down{\anc{p}{y}} \} \\
  \ind \ind \ind DO \\
  \ind \ind \ind \ind $y := \aget{p}{y}$ \\
  \ind \ind \ind OD \\
  \ind \ind [ find\_set\_postcondition $p$ $x$ $y$ ]'' \\
  \ind \textbf{apply} vcg\_tc\_simp \\
  \ind -- proof of three verification conditions omitted
\end{numquote}
In line 4, variable $y$ is initialised with the start node $x$.
The while-loop stops when it finds a node that is its own parent in line 5.
Otherwise it continues with the parent of the current node in line 9.

\vspace{1.8mm}
The precondition requires that $p$ is a forest (representing the disjoint sets) and $x$ is a point (representing a node in the forest):
\begin{quote}
  find\_set\_precondition $p$ $x$ $\iff$ forest $p$ $\wedge$ point $x$
\end{quote}
The loop invariant requires the precondition and that $y$ is a point reachable from $x$ along a chain of parents:
\begin{quote}
  find\_set\_invariant $p$ $x$ $y$ $\iff$ find\_set\_precondition $p$ $x$ $\wedge$ point $y$ $\wedge$ $y \lleq \anc{p}{x}$
\end{quote}
Vector $\anc{p}{x} = \cnv{p}\astop \comp x$ describes the ancestors of $x$ under $p$ using array-read notation, that is, all successors of $x$ under zero or more transitions of $p$.
The postcondition states that $y$ is a point and the root of the tree containing $x$:
\begin{quote}
  find\_set\_postcondition $p$ $x$ $y$ $\iff$ point $y$ $\wedge$ $y$ = \kroot{p}{x}
\end{quote}
A vector of all roots of the disjoint-set forest represented by $p$ is constructed from the relation $p \meet \unit$ containing all self-loops of $p$:
\begin{quote}
  $\kroots{p} = (p \meet \unit) \comp \ltop$
\end{quote}
\eject
The root of a node $x$ in forest $p$ is the unique root that is an ancestor of $x$:
\begin{quote}
  $\kroot{p}{x} = \anc{p}{x} \meet \kroots{p}$
\end{quote}
Part 1 of the following result gives an equivalent characterisation.
Part 2 shows that following the parents of roots one or several times gives the roots again.
We discuss a constructive proof of part 3 at the end of Section \ref{section.find-set}.

\begin{thm}
  \label{theorem.root}
  \begin{enumerate}
  \item[]
  \item $\kroot{p}{x} = (p \meet \unit) \comp \anc{p}{x}$.\afplemma{root_var}
  \item $\kroot{p}{x} = \aread{p}{\kroot{p}{x}} = \anc{p}{\kroot{p}{x}}$ and $\kroots{p} = \aread{p}{\kroots{p}} = \anc{p}{\kroots{p}}$ if $p$ is univalent.\afplemma{root_successor_loop}%
        \label{theorem.root.roots-successor-loop}
  \item $\kroot{p}{x}$ is a point if $p$ is a forest and $x$ is a point.\afplemma{root_point}
        \label{theorem.root.point}
  \end{enumerate}
\end{thm}

For partial correctness, the three verification conditions generated for find-set are:
\begin{enumerate}
\item find\_set\_precondition $p$ $x$ $\implies$ find\_set\_invariant $p$ $x$ $x$
\item find\_set\_invariant $p$ $x$ $y$ $\wedge$ $y \neq \aget{p}{y}$ $\implies$ find\_set\_invariant $p$ $x$ $\aget{p}{y}$
\item find\_set\_invariant $p$ $x$ $y$ $\wedge$ $y = \aget{p}{y}$ $\implies$ find\_set\_postcondition $p$ $x$ $y$
\end{enumerate}
To maintain the invariant we can assume that the condition of the while-loop holds.
To show the postcondition we can assume that the condition of the while-loop does not hold.
For total correctness, the first and third verification conditions are the same but maintenance of the invariant is modified taking into account the variant of the while-loop:
\begin{enumerate}
\setcounter{enumi}{1}
\item find\_set\_invariant $p$ $x$ $y$ $\wedge$ $y \neq \aget{p}{y}$ $\wedge$ \down{\anc{p}{y}} = $n$ $\implies$
      find\_set\_invariant $p$ $x$ $\aget{p}{y}$ $\wedge$ \down{\anc{p}{\aread{p}{y}}} $<$ $n$
\end{enumerate}
The variable $n$ transports the initial value of the variant from the assumption to the conclusion, where it is compared with the final value of the variant.
For the above program, the variant is \down{\anc{p}{y}}, the number of elements in the algebra below $\anc{p}{y}$.
The expression $\anc{p}{y}$ is a vector representing the ancestors of $y$.
The variant is an order-preserving expression that turns this vector into a natural number.

See Appendix \ref{section.proofs} for a proof of the total-correctness verification conditions generated for find-set.

\medskip
We now discuss Theorem \ref{theorem.root}.\ref{theorem.root.point}, which states that the root of the tree containing point $x$ in the forest $p$ is a point, that is, a vector representing a single node.
This result could be proved by working with the definitions of roots, points and forests.
We give a different proof based on find-set.
Observe that this operation computes the desired root and the postcondition states it is a point.
Moreover the precondition of find-set contains just the assumptions of Theorem \ref{theorem.root}.\ref{theorem.root.point}.
Hence this result immediately follows from total correctness of find-set.
In Isabelle/HOL, similary to make-set discussed in Section \ref{section.make-set} we obtain:
\begin{flushleft}
  \textbf{lemma} find\_set\_exists: ``find\_set\_precondition $p$ $x$ $\implies$ $\exists y$ . find\_set\_postcondition $p$ $x$ $y$'' \\
  \ind \textbf{using} tc\_extract\_function find\_set \textbf{by} blast
\end{flushleft}

\eject
Theorem \ref{theorem.root}.\ref{theorem.root.point} then is a simple consequence:
\begin{flushleft}
  \textbf{lemma} root\_point: ``forest $p$ $\wedge$ point $x$ $\implies$ point (\kroot{p}{x})'' \\
  \ind \textbf{using} find\_set\_exists find\_set\_precondition\_def find\_set\_postcondition\_def \textbf{by} simp
\end{flushleft}
Essentially this is a constructive proof using the imperative programs supported by the Hoare-logic library.
This method does not necessarily reduce the amount of work needed for proving a result but shifts the work to the correctness proof of a program.
However, once the correctness proof is established it saves additional work.
Moreover, this approach facilitates computational reasoning.

\subsection{Path compression and its effect}
\label{section.path-compression}

Path compression is a technique to decrease the depth of disjoint-set forests, which makes subsequent find-set operations faster.
The idea is to change the parent of every node encountered during the execution of find-set to the root of the tree.
Because the root is known only after the chain of parents has been traversed, modifying the parents takes place in a separate traversal.
In a recursive implementation of find-set, these modifications would take place on the way out from the recursion.
We use two while-loops for the same purpose.
The first loop is the find-set operation described in Section \ref{section.find-set} to find the root $y$ of the tree.
As shown here, the second loop traverses the same sequence of nodes and adjusts the parents on the way:
\begin{numquote}
  \textbf{theorem} path\_compression:\afptheorem{path_compression} \\
  \ind ``VARS $p$ $t$ $w$ \\
  \ind \ind [ path\_compression\_precondition $p$ $x$ $y$ $\wedge$ $p_0$ = $p$ ] \\
  \ind \ind $w := x$; \\
  \ind \ind WHILE $y \neq \aget{p}{w}$ \\
  \ind \ind \ind INV \{ path\_compression\_invariant $p$ $x$ $y$ $p_0$ $w$ \} \\
  \ind \ind \ind VAR \{ \down{\anc{p}{w}} \} \\
  \ind \ind \ind DO \\
  \ind \ind \ind \ind $t := w$; \\
  \ind \ind \ind \ind $w := \aget{p}{w}$; \\
  \ind \ind \ind \ind $\aset{p}{t}{y}$ \\
  \ind \ind \ind OD \\
  \ind \ind [ path\_compression\_postcondition $p$ $x$ $y$ $p_0$ ]'' \\
  \ind \textbf{apply} vcg\_tc\_simp \\
  \ind -- proof of three verification conditions omitted
\end{numquote}
This program is executed immediately after the while-loop of find-set, where $p$ is the parent array, $x$ is the original node and $y$ is its representative computed by find-set, which is the root of the tree that contains $x$.
The assignments in lines 4 and 10 traverse the same sequence of nodes as find-set, except according to line 5 this loop finishes immediately before reaching the root.
Lines 9 and 11 set the parent of the current node to the root.
Temporary variable $t$ is used to save the current node $w$, which is changed by line 10.

\medskip
The variant in line 7 is the same as the one used for find-set, except the current node is now stored in $w$.
Also the generated verification conditions have the same structure as in the proof for find-set.
It remains to discuss the actual precondition, invariant and postcondition.
The precondition is:
\begin{quote}
  path\_compression\_precondition $p$ $x$ $y$ $\iff$ forest $p$ $\wedge$ point $x$ $\wedge$ point $y$ $\wedge$ $y$ = \kroot{p}{x}
\end{quote}
It extends the precondition of find-set by two conditions, which are just the postcondition of find-set.
This ensures the two loops can be composed sequentially.
The invariant significantly extends the precondition:
\begin{quote}
  path\_compression\_invariant $p$ $x$ $y$ $p_0$ $w$ $\iff$ \\
  \ind path\_compression\_precondition $p$ $x$ $y$ $\wedge$ forest $p_0$ $\wedge$ $\fc(p)$ = $\fc(p_0)$ $\wedge$ $p \meet \unit = p_0 \meet \unit$ \\
  \ind $\wedge$ point $w$ $\wedge$ $w \lleq \anc{p_0}{x}$ $\wedge$ $y$ = \kroot{p}{w} $\wedge$ $\awrite{p_0}{(\anc{p_0}{x} - \anc{p_0}{w})}{y} = p$
\end{quote}
First, the invariant requires that $p_0$ is a forest.
Second, $\fc(p)$ = $\fc(p_0)$ states that the components of $p$ do not change, that is, $p$ represents the same disjoint sets.
Third, $p \meet \unit = p_0 \meet \unit$ states that the roots of the component trees of $p$ do not change.
Next, node $w$ is an ancestor of $x$ in $p_0$, and $y$ is the representative in $p$ of the set containing $w$.

\medskip
The final condition $\awrite{p_0}{(\anc{p_0}{x} - \anc{p_0}{w})}{y} = p$ shows the effect of path compression while the loop is in progress.
This specification contains an update of array $p_0$ at all nodes that are in vector $\anc{p_0}{x} - \anc{p_0}{w}$.
Hence, the parent array is obtained from the original parent array $p_0$ by updating to value $y$ the entries for nodes that are ancestors of $x$ but not ancestors of $w$.
These are just the nodes between $x$ inclusively and $w$ exclusively.

\medskip
The postcondition is part of the invariant:
\begin{quote}
  path\_compression\_postcondition $p$ $x$ $y$ $p_0$ $\iff$ \\
  \ind forest $p$ $\wedge$ $y$ = \kroot{p}{x} $\wedge$ $\fc(p)$ = $\fc(p_0)$ $\wedge$ $p \meet \unit = p_0 \meet \unit$ $\wedge$ $\awrite{p_0}{\anc{p_0}{x}}{y} = p$
\end{quote}
For correctness we only require that path compression does not change the disjoint sets represented by the forest.
We also get that the roots do not change.
The final condition $\awrite{p_0}{\anc{p_0}{x}}{y} = p$ expresses the precise effect of path compression: the resulting parent array $p$ is obtained from $p_0$ by updating the entries for ancestors of $x$ to the value $y$.

\medskip
Appendix \ref{section.forest-properties} discusses a selection of results used for maintaining the invariant; see Theorem \ref{theorem.path-compression}.

In Section \ref{section.init-sets} we showed how disjoint-set forests can be initialised either by a loop applying make-set to elements one at a time or by a single assignment to the parent array.
We now discuss an implementation of path compression similarly by one assignment.

\medskip
Path compression updates the parent array at all nodes on the path from the start $x$ to the root $y$ of the same tree.
The set of updated nodes therefore comprises all ancestors of $x$, obtained by the vector $\anc{p}{x}$.
Moreover, each update assigns the same value $y$ to the parent.
This is also evident from the postcondition of path compression.
Hence the effect of path compression is obtained by the single assignment:
\begin{quote}
  $\aset{p}{\anc{p}{x}}{y}$
\end{quote}
The following result confirms this:
\begin{numquote}
  \textbf{theorem} path\_compression\_assign:\afptheorem{path_compression_assign} \\
  \ind ``VARS $p$ \\
  \ind \ind [ path\_compression\_precondition $p$ $x$ $y$ $\wedge$ $p_0$ = $p$ ] \\
  \ind \ind $\aset{p}{\anc{p}{x}}{y}$ \\
  \ind \ind [ path\_compression\_postcondition $p$ $x$ $y$ $p_0$ ]'' \\
  \ind \textbf{apply} vcg\_tc\_simp \\
  \ind -- proof of one verification condition omitted
\end{numquote}
Such relational assignments can be used to obtain higher-level, more concise implementations.
Moreover the underlying array updates can be used in specifications as the postcondition of path compression demonstrates.

\subsection{The find-set operation with path compression and its effect}
\label{section.find-set-path-compression}

Using the technique of Section \ref{section.make-set} we extract function definitions for the find-set operation of Section \ref{section.find-set} and the path-compression operation of Section \ref{section.path-compression}.
This allows us to combine the two programs into the following one with a simple correctness proof:
\begin{numquote}
  \textbf{theorem} find\_set\_path\_compression:\afptheorem{find_set_path_compression} \\
  \ind ``VARS $p$ $y$ \\
  \ind \ind [ find\_set\_precondition $p$ $x$ $\wedge$ $p_0$ = $p$ ] \\
  \ind \ind $y$ := find\_set $p$ $x$; \\
  \ind \ind $p$ := path\_compression $p$ $x$ $y$ \\
  \ind \ind [ path\_compression\_postcondition $p$ $x$ $y$ $p_0$ ]'' \\
  \ind \textbf{apply} vcg\_tc\_simp \\
  \ind \textbf{using} find\_set\_function find\_set\_postcondition\_def find\_set\_precondition\_def \\
  \ind \ind path\_compression\_function path\_compression\_precondition\_def \textbf{by} fastforce
\end{numquote}
We can also extract a function for this program, but this function returns a pair of values as the find-set operation with path compression both modifies the disjoint-set forest and returns the root of the tree containing node $x$:
\begin{flushleft}
  \textbf{definition} ``find\_set\_path\_compression $p$ $x$ = \\
  \ind (SOME $(p',y)$ . path\_compression\_postcondition $p'$ $x$ $y$ $p$)''
\end{flushleft}
The function extracted for find\_set\_path\_compression returns values which satisfy the postcondition of path\_compression.
It no longer contains the information that the return value was constructed using find\_set and path\_compression.
In other words, part of the connection between the function extracted for find\_set\_path\_compression and the composition of the functions extracted for find\_set and path\_compression is lost.

\medskip
We can prove that the root of the tree returned by find\_set\_path\_compression is the result of find\_set.
We can also prove that the forest returned by find\_set\_path\_compression has the same roots as the forest returned by the composition of find\_set and path\_compression.
Moreover we can prove that the two forests have the same semantics, that is, represent the same equivalence relation.
To show that the two forests are equal, however, it is necessary to know precisely how path compression modifies the parent array, which is captured by the postcondition of path\_compression given in Section \ref{section.path-compression}.

\subsection{Path splitting}
\label{section.path-splitting}

As shown in Section \ref{section.find-set-path-compression}, finding the representative with path compression takes two iterations.
Alternative path-compression techniques have been developed which require only one iteration and may perform better in practice \cite{LeeuwenWeide1977,TarjanLeeuwen1984}.
We discuss two such techniques in this and the next section.

\medskip
Path splitting makes one iteration from the start node to the root of its tree and updates the parent array to point to grandparents along the way.
This shortens paths towards the root for subsequent find-set operations:
\begin{numquote}
  \textbf{theorem} find\_path\_splitting:\afptheorem{find_path_splitting} \\
  \ind ``VARS $p$ $t$ $y$ \\
  \ind \ind [ find\_set\_precondition $p$ $x$ $\wedge$ $p_0$ = $p$ ] \\
  \ind \ind $y := x$; \\
  \ind \ind WHILE $y \neq \aget{p}{y}$ \\
  \ind \ind \ind INV \{ path\_splitting\_invariant $p$ $x$ $y$ $p_0$ \} \\
  \ind \ind \ind VAR \{ \down{\anc{p}{y}} \} \\
  \ind \ind \ind DO \\
  \ind \ind \ind \ind $t := \aget{p}{y}$; \\
  \ind \ind \ind \ind $\aset{p}{y}{\aget{p}{\aget{p}{y}}}$; \\
  \ind \ind \ind \ind $y := t$ \\
  \ind \ind \ind OD \\
  \ind \ind [ path\_splitting\_postcondition $p$ $x$ $y$ $p_0$ ]'' \\
  \ind \textbf{apply} vcg\_tc\_simp \\
  \ind -- proof of three verification conditions omitted
\end{numquote}
The program is similar to find\_set except the additional compression update in line 10, which makes it necessary to use temporary variable $t$ to store the parent of $y$ in line 9.
The verification condition generator with simplification eliminates the temporary variable so the condition to maintain the invariant is the same that would be obtained for the simultaneous assignment:
\begin{quote}
  $\aset{y, p}{y}{\aget{p}{y}, \aget{p}{\aget{p}{y}}}$
\end{quote}
The pre- and postconditions of find\_path\_splitting are the same as those of find\_set\_path\_compression except the effect of path splitting is different:
\begin{quote}
  path\_splitting\_postcondition $p$ $x$ $y$ $p_0$ $\iff$ \\
  \ind forest $p$ $\wedge$ $y$ = \kroot{p}{x} $\wedge$ $\fc(p) = \fc(p_0)$ $\wedge$ $p \meet \unit = p_0 \meet \unit$ $\wedge$ $\awrite{p_0}{\anc{p_0}{x}}{\cnv{(p_0 \comp p_0)}} = p$
\end{quote}
The update $\awrite{p_0}{\anc{p_0}{x}}{\cnv{(p_0 \comp p_0)}} = p$ specifies that all ancestors of $x$ are updated according to the grandparent relation $p_0 \comp p_0$.
Hence also the effect of path splitting can be expressed as a single assignment to the parent array.

\eject
\noindent The postcondition can be proved using the loop invariant:
\begin{quote}
  path\_splitting\_invariant $p$ $x$ $y$ $p_0$ $\iff$ \\
  \ind find\_set\_precondition $p$ $x$ $\wedge$ point $y$ $\wedge$ $y \lleq \anc{p_0}{x}$ $\wedge$ forest $p_0$ \\[1mm]
  \ind $\wedge$ $\awrite{p_0}{(\anc{p_0}{x} - \anc{p_0}{y})}{\cnv{(p_0 \comp p_0)}} = p$
\end{quote}
Here again, while the loop is in progress only the nodes between $x$ inclusively and $y$ exclusively have been updated, this time to point to their grandparents.
A number of useful properties, including some which were part of the invariant of path\_compression, follow from path\_splitting\_invariant:
\begin{quote}
  path\_splitting\_invariant $p$ $x$ $y$ $p_0$ $\implies$ \\
  \ind $\aget{p}{y} = \aget{p_0}{y}$ $\wedge$ $\aget{p}{\aget{p}{y}} = \aget{p_0}{\aget{p_0}{y}}$ $\wedge$ $\fc(p) = \fc(p_0)$ $\wedge$ $p \meet \unit = p_0 \meet \unit$
\end{quote}
This is because the invariant reflects the precise effect of path splitting.

\subsection{Path halving}
\label{section.path-halving}

Path halving is similar to path splitting except the parent of only every second node is updated to the grandparent:
\begin{numquote}
  \textbf{theorem} find\_path\_halving:\afptheorem{find_path_halving} \\
  \ind ``VARS $p$ $y$ \\
  \ind \ind [ find\_set\_precondition $p$ $x$ $\wedge$ $p_0$ = $p$ ] \\
  \ind \ind $y := x$; \\
  \ind \ind WHILE $y \neq \aget{p}{y}$ \\
  \ind \ind \ind INV \{ path\_halving\_invariant $p$ $x$ $y$ $p_0$ \} \\
  \ind \ind \ind VAR \{ \down{\anc{p}{y}} \} \\
  \ind \ind \ind DO \\
  \ind \ind \ind \ind $\aset{p}{y}{\aget{p}{\aget{p}{y}}}$; \\
  \ind \ind \ind \ind $y := \aget{p}{y}$ \\
  \ind \ind \ind OD \\
  \ind \ind [ path\_halving\_postcondition $p$ $x$ $y$ $p_0$ ]'' \\
  \ind \textbf{apply} vcg\_tc\_simp \\
  \ind -- proof of three verification conditions omitted
\end{numquote}
The updates in lines 9--10 amount to the simultaneous assignment:
\begin{quote}
  $\aset{y, p}{y}{\aget{p}{\aget{p}{y}}, \aget{p}{\aget{p}{y}}}$
\end{quote}
It shows that the loop proceeds from each node to its grandparent.
This is reflected in the postcondition:
\begin{quote}
  path\_halving\_postcondition $p$ $x$ $y$ $p_0$ $\iff$ \\
  \ind forest $p$ $\wedge$ $y$ = \kroot{p}{x} $\wedge$ $\fc(p) = \fc(p_0)$ $\wedge$ $p \meet \unit = p_0 \meet \unit$ \\
  \ind $\wedge$ $\awrite{p_0}{\anc{(p_0 \comp p_0)}{x}}{\cnv{(p_0 \comp p_0)}} = p$
\end{quote}
Vector $\anc{(p_0 \comp p_0)}{x}$ contains the nodes reachable in an even number of steps from $x$.
Again this means that the effect of path halving can be expressed as a single assignment to the parent array.
The following invariant is used to prove the postcondition:
\begin{quote}
  path\_halving\_invariant $p$ $x$ $y$ $p_0$ $\iff$ \\
  \ind find\_set\_precondition $p$ $x$ $\wedge$ point $y$ $\wedge$ $y \lleq \anc{p_0}{x}$ $\wedge$ forest $p_0$ \\[1mm]
  \ind $\wedge$ $\awrite{p_0}{(\anc{(p_0 \comp p_0)}{x} - \anc{p_0}{y})}{\cnv{(p_0 \comp p_0)}} = p$
\end{quote}
Vector $\anc{(p_0 \comp p_0)}{x} - \anc{p_0}{y}$ contains all nodes which are reachable from $x$ in an even number of steps but not an ancestor of $y$.
This is every second node between $x$ inclusively and $y$ exclusively.
Again the precise knowledge about the effect of path halving allows us to derive useful properties so they do not have to be included in the invariant:
\begin{quote}
  path\_halving\_invariant $p$ $x$ $y$ $p_0$ $\implies$ \\
  \ind $\aget{p}{y} = \aget{p_0}{y}$ $\wedge$ $\aget{p}{\aget{p}{y}} = \aget{p_0}{\aget{p_0}{y}}$ $\wedge$ $\fc(p) = \fc(p_0)$ $\wedge$ $p \meet \unit = p_0 \meet \unit$
\end{quote}
Appendix \ref{section.forest-properties} discusses properties used for verifying the correctness of path splitting and path halving; see Theorem \ref{theorem.square}.

\section{Verifying disjoint-set forests -- union and ranks}
\label{section.verification-union}

In this section we implement the union-sets operation of disjoint-set forests and the union-by-rank strategy and verify correctness of these algorithms.
Section \ref{section.union-sets} covers the basic union-sets operation without ranks, and recalls the concept of ranks.
Peano structures used to represent numbers are studied in Section \ref{section.peano}.
Section \ref{section.init-ranks} covers the initialisation of ranks.
Union-by-rank is implemented and verified in Section \ref{section.union-by-rank}.

\subsection{The union-sets operation}
\label{section.union-sets}

We start by considering the union-sets operation, which takes two elements and joins the corresponding disjoint sets into a single set.
To this end it finds the representatives of the equivalence classes of the elements and links one to the other:
\begin{numquote}
  \textbf{theorem} union\_sets\_2:\afptheorem{union_sets_2} \\
  \ind ``VARS $p$ $r$ $s$ \\
  \ind \ind [ union\_sets\_precondition $p$ $x$ $y$ $\wedge$ $p_0$ = $p$ ] \\
  \ind \ind $r$ := find\_set $p$ $x$; \\
  \ind \ind $p$ := path\_compression $p$ $x$ $r$; \\
  \ind \ind $s$ := find\_set $p$ $y$; \\
  \ind \ind $p$ := path\_compression $p$ $y$ $s$; \\
  \ind \ind $\aset{p}{r}{s}$ \\
  \ind \ind [ union\_sets\_postcondition $p$ $x$ $y$ $p_0$ ]'' \\
  \ind \textbf{apply} vcg\_tc\_simp \\
  \ind -- proof of one verification condition omitted
\end{numquote}
We have also verified an alternative implementation using the combined find\_set\_path\_compression of Section \ref{section.find-set-path-compression}.
Note how the forest $p$ is threaded through both occurrences of path\_compression, where it may be modified, before line 8 adds the link from the root $r$ of the tree containing $x$ to the root $s$ of the tree containing $y$.

\medskip
The precondition of union-sets requires that $p$ represents a forest and $x$ and $y$ represent single nodes:
\begin{quote}
  union\_sets\_precondition $p$ $x$ $y$ $\iff$ forest $p$ $\wedge$ point $x$ $\wedge$ point $y$
\end{quote}
The postcondition also requires that the final value of $p$ represents the equivalence relation where the sets containing $x$ and $y$ have been merged into one:
\begin{quote}
  union\_sets\_postcondition $p$ $x$ $y$ $p_0$ $\iff$ forest $p$ $\wedge$ $\fc(p)$ = $\wcc(p_0 \join (x \comp \cnv{y}))$
\end{quote}
To get the latter equivalence relation, we add the pair $(x,y)$ to the initial equivalence relation $p_0$ and compute its equivalence closure.
The pair $(x,y)$ is described by $x \comp \cnv{y}$ since $x$ and $y$ are points.

Appendix \ref{section.forest-properties} discusses a selection of results used for proving the correctness union-sets; see Theorem \ref{theorem.union-sets}.

Correctness of the union-sets operation would not be affected if the assignment in line 8 was replaced with $\aset{p}{s}{r}$.
More efficient implementations of union-sets therefore decide which of these two assignments to use based on heuristics such as union-by-rank \cite{TarjanLeeuwen1984}.
The rank of a node is a natural number giving an upper bound on the depth of the subtree at the node.
It is more efficient to add a link from the root with smaller rank to the other.
Using ranks in a disjoint-set forest implementation requires comparisons and simple arithmetic operations.
We implement this extension in the relation-algebraic framework using basic Peano arithmetic.

\subsection{Peano structures}
\label{section.peano}

Disjoint-set forests and their algorithms can be implemented using many different data structures in Isabelle/HOL; for example, \cite{LammichMeis2012} uses lists and arrays over the built-in type of natural numbers.
The aim of this paper is to verify the correctness of disjoint-set forests using relation-algebraic structures, hence we need to encode natural numbers as elements of relation algebras.
Peano arithmetic offers a simple yet suitable foundation for such an encoding.

Peano arithmetic describes the natural numbers based on the constant $\opzero$ and a successor operation $\opsucc$.
In a relational setting, numbers are represented as points.
A point $\zero$ represents the number $\opzero$.
An injective mapping $\suc$ represents the successor operation $\opsucc$.
The successor of a number represented by point $x$ is obtained by the relational operation $\cnv{\suc} \comp x$.
The composition $\suc \comp x$ results in the predecessor of $x$.
Our axiomatisation of Peano structures using relation algebras is based on \cite{BerghammerZierer1986,Berghammer2020}.

\medskip
The following lists the arithmetic operations we use for union-by-rank:
\begin{quote}
  $\opzero = \zero$ \\
  $\opsucc(x) = \cnv{\psuc} \comp x$ \\
  $x < y \iff x \lleq \psuc^+ \comp y$
\end{quote}
For technical reasons discussed below we use a modified successor operation $\psuc = \suc - \cnv{\zero}$ which also works for finite sets of numbers $\{ \opzero, \dots, m-1 \}$.
The rest of this section studies properties of this relational encoding of numbers and discusses some challenges of formalising it in Isabelle/HOL.

Ideally, the rank information of nodes would be represented by a heterogeneous relation between the set of nodes and the set of natural numbers.
However, the class hierarchy for semirings, lattices, relation algebras, Kleene algebras and similar structures in Isabelle/HOL models only homogeneous relations.
A formalisation using set-theoretic functions and heterogeneous relations would be possible, but this would not support the use of algebras for verification because no formalisation of heterogeneous relation algebras is available in Isabelle/HOL.

Moreover the set of nodes in disjoint-set forests is finite (which ensures, for example, that initialisation of the forest terminates).
Hence we have to use a finite subset of the natural numbers which has as many elements as there are nodes in the disjoint-set forest.
Since the ranks in a forest with $m$ nodes will be at most $m-1$, it is enough to use the numbers $\opzero, \dots, m-1$.

We therefore axiomatise the zero constant $\zero$ and the partial successor relation $\psuc$ that maps each number except $m-1$ to its successor.
This is achieved in two steps.
We first consider the total successor relation $\suc$ that maps each number to its successor modulo $m$.
The desired partial successor relation $\psuc$ is derived from it in the second step.

\medskip
For the first step we simply use a subset of the Peano axioms:
\begin{enumerate}
\item point $\zero$
\item mapping $\suc$
\item injective $\suc$
\item $\cnv{\suc}\astop \comp \zero = \ltop$
\end{enumerate}
Axiom 1 specifies that $\opzero$ is a number.
Axiom 2 specifies that every number has exactly one successor.
Axiom 3 specifies that numbers with the same successor are equal.
Axiom 4 specifies that every number can be obtained from $\opzero$ by finitely many applications of the successor.
We omit the Peano axiom $\suc \comp \zero = \lbot$ which would specify that $\opzero$ is not the successor of any number.

Since this is a weakening of the Peano axioms, the natural numbers are a model, but so are the numbers modulo $m$.
The following result gives consequences of the above four axioms.

\begin{thm}
  \label{theorem.z-s}
  \begin{enumerate}
  \item[]
  \item $\suc \meet \unit \lleq \zero$.\afplemma{S_inf_1_below_Z}
        \label{theorem.z-s.s-inf-1-below-z}
  \item $\zero \lleq \suc^*$.\afplemma{Z_below_S_star}
  \item $\cnv{\suc}\astop \comp \suc^* = \ltop$.\afplemma{S_connected}
  \item $\suc^* \join \cnv{\suc}\astop = \ltop$.\afplemma{S_star_connex}
  \item $\zero \join (\cnv{\suc} \comp \ltop) = \ltop$.\afplemma{Z_sup_conv_S_top}
        \label{theorem.z-s.top-s-sup-conv-z}
  \end{enumerate}
\end{thm}

\begin{prf}
  See Appendix \ref{section.proofs} for proofs of Theorems \ref{theorem.z-s}.\ref{theorem.z-s.s-inf-1-below-z} and \ref{theorem.z-s}.\ref{theorem.z-s.top-s-sup-conv-z}, and \cite{Guttmann2020c} for all proofs.
\end{prf}

Part 4 shows that the preorder $\suc^*$ is connex, that is, any two numbers are comparable.
Part 5 shows that every number is $\opzero$ or the successor of a number.

The successor operation provides a convenient way to compare two natural numbers.
Namely, $x \leq y$ if $y$ can be reached from $x$ by finitely many applications of the successor, formally $y \lleq \cnv{\suc}\astop \comp x$ or equivalently $x \lleq \suc^* \comp y$ for points $x$ and $y$.
This does not work for numbers modulo $m$ since comparison depends on the chosen representative.

In the second step, we therefore derive a partial successor relation $\psuc = \suc - \cnv{\zero}$, which computes the successor for all numbers except $m-1$.
The modified $\psuc$ is obtained from $\suc$ by removing any entries that map to $\opzero$; in the intended model this is just the mapping from $m-1$ to $\opzero$.

We have verified the soundness of the relational Peano axioms in Isabelle/HOL.
Specifically, we have defined a model of natural numbers modulo $m$ as relations and proved that this model satisfies the above axioms for $\zero$ and $\suc$.
We have also shown that the natural numbers below $m$, taken from Isabelle/HOL's nat type, order-embed into our relational model.
The order-embedding preserves the arithmetic operations $0$, partial successor and the $<$ relation on this subset of nat.
Moreover, the relational model satisfies the axioms of all algebraic structures used for the results of this paper.
See the Isabelle/HOL theories for further details about this model and proofs of these results.

The following result shows that $\psuc$ satisfies most of the properties of $\suc$ and some additional properties.

\begin{thm}
  \label{theorem.s'}
  \begin{enumerate}
  \item[]
  \item $\psuc$ is univalent.\afplemma{S'_univalent}
  \item $\psuc$ is injective.\afplemma{S'_injective}
  \item $\cnv{\psuc}\astop \comp \zero = \ltop$.\afplemma{S'_star_Z_top}
        \label{theorem.s'.s'-star-z-top}
  \item $\psuc$ is irreflexive.\afplemma{S'_irreflexive}
  \item $\zero \lleq \psuc^*$.\afplemma{Z_below_S'_star}
  \item $\cnv{\psuc}\astop \comp \psuc^* = \ltop$.\afplemma{S'_connected}
  \item $\psuc^* \join \cnv{\psuc}\astop = \ltop$.\afplemma{S'_star_connex}
  \item $\zero \join (\cnv{\psuc} \comp \ltop) = \ltop$.\afplemma{Z_sup_conv_S'_top}
  \item $\psuc \comp \zero = \lbot$.\afplemma{S'_Z}
  \end{enumerate}
\end{thm}

\begin{prf}
  See Appendix \ref{section.proofs} for a proof of Theorem \ref{theorem.s'}.\ref{theorem.s'.s'-star-z-top} and \cite{Guttmann2020c} for all proofs.
\end{prf}

Part 9 shows that $\opzero$ has no predecessor under $\psuc$.
We next consider the predecessor of $\opzero$ under $\suc$ given by the relation $\maxi = \suc \comp \zero$.
The following result characterises the two models: the natural numbers and the natural numbers modulo $m$.

\begin{thm}
  \label{theorem.m}
  \begin{enumerate}
  \item[]
  \item $\maxi$ is a point if and only if $\suc$ is surjective.\afplemma{M_point_iff_S_surjective}
        \label{theorem.m.m-point-iff-s-surjective}
  \item $\maxi \neq \lbot$ if and only if $\suc$ is surjective.\afplemma{M_bot_iff_S_not_surjective}
        \label{theorem.m.m-bot-iff-s-not-surjective}
  \item $\maxi = \lbot$ or $\maxi$ is a point.\afplemma{M_point_or_bot}
        \label{theorem.m.m-point-or-bot}
  \end{enumerate}
\end{thm}

\begin{prf}
  See Appendix \ref{section.proofs} or \cite{Guttmann2020c} for a proof of Theorem \ref{theorem.m}.
\end{prf}

In the natural-number model, $\opzero$ has no predecessor under $\suc$ so $\suc$ is not surjective; then $\maxi = \lbot$.
In the modulo-$m$ model, $\opzero$ has exactly one predecessor under $\suc$ and $\suc$ is bijective; then $\maxi$ is a point.

The relation $\maxi$ facilitates an alternative characterisation of $\psuc$ as the following result shows.
We also characterise the model $m=1$, that is, $\maxi = \zero$.

\begin{thm}
  \label{theorem.m-z}
  \begin{enumerate}
  \item[]
  \item $\psuc = \suc - \maxi$.\afplemma{S'_var}
        \label{theorem.m-z.s'-var}
  \item $\maxi = \zero$ if and only if $\unit = \ltop$.\afplemma{M_is_Z_iff_1_is_top}
  \item $\suc$ is irreflexive if $\maxi \neq \zero$.\afplemma{S_irreflexive}
  \end{enumerate}
\end{thm}

\begin{prf}
  See Appendix \ref{section.proofs} for a proof of Theorem \ref{theorem.m-z}.\ref{theorem.m-z.s'-var} and \cite{Guttmann2020c} for all proofs.
\end{prf}

According to part 1, the partial $\psuc$ can also be obtained from $\suc$ by removing the entry in row $m-1$.
Part 3 shows that $\suc$ is irreflexive unless $m=1$.

\subsection{Initialising ranks}
\label{section.init-ranks}

For the union-by-rank strategy to work, the rank of each element needs to be initialised to $\opzero$.
To this end we augment init-sets from Section \ref{section.init-sets} by adding the assignment in line 9:
\begin{numquote}
  \textbf{theorem} init\_ranks:\afptheorem{init_ranks} \\
  \ind ``VARS $h$ $p$ $x$ $\rank$ \\
  \ind \ind [ True ] \\
  \ind \ind FOREACH $x$ \\
  \ind \ind \ind USING $h$ \\
  \ind \ind \ind INV \{ $p - h = \unit - h$ $\wedge$ $\rank - h = \cnv{\zero} - h$ \} \\
  \ind \ind \ind DO \\
  \ind \ind \ind \ind $p$ := make\_set $p$ $x$; \\
  \ind \ind \ind \ind $\aset{\rank}{x}{\opzero}$ \\
  \ind \ind \ind OD \\
  \ind \ind [ $p = \unit$ $\wedge$ forest $p$ $\wedge$ $h = \lbot \wedge \rank = \cnv{\zero}$ $\wedge$ rank\_property $p$ $\rank$ ]'' \\
  \ind \textbf{apply} vcg\_tc\_simp \\
  \ind -- proof of one verification condition omitted
\end{numquote}

The overall effect is to set $\rank := \cnv{\zero}$ in addition to setting $p := \unit$.
The loop invariant reflects the partial state of these updates.
The postcondition establishes additional properties of the rank array which are discussed below.

\subsection{Union-by-rank}
\label{section.union-by-rank}

The implementation of union-by-rank extends the algorithm presented in Section \ref{section.union-sets}:
\begin{numquote}\small
  \textbf{theorem} union\_sets\_by\_rank:\afptheorem{union_sets_by_rank} \\
  \ind ``VARS $p$ $r$ $s$ $\rank$ \\
  \ind \ind [ union\_sets\_precondition $p$ $x$ $y$ $\wedge$ $p_0 = p$ $\wedge$ rank\_property $p$ $\rank$ ] \\
  \ind \ind $r$ := find\_set $p$ $x$; \\
  \ind \ind $p$ := path\_compression $p$ $x$ $r$; \\
  \ind \ind $s$ := find\_set $p$ $y$; \\
  \ind \ind $p$ := path\_compression $p$ $y$ $s$; \\
  \ind \ind IF $r \neq s$ THEN \\
  \ind \ind \ind IF $\aget{\rank}{r} < \aget{\rank}{s}$ THEN \\
  \ind \ind \ind \ind $\aset{p}{r}{s}$ \\
  \ind \ind \ind ELSE \\
  \ind \ind \ind \ind $\aset{p}{s}{r}$; \\
  \ind \ind \ind \ind IF $\aget{\rank}{r} = \aget{\rank}{s}$ THEN \\
  \ind \ind \ind \ind \ind $\aset{\rank}{r}{\opsucc(\aget{\rank}{r})}$ \\
  \ind \ind \ind \ind FI \\
  \ind \ind \ind FI \\
  \ind \ind FI \\
  \ind \ind [ union\_sets\_postcondition $p$ $x$ $y$ $p_0$ $\wedge$ rank\_property $p$ $\rank$ ]'' \\
  \ind \textbf{apply} vcg\_tc\_simp \\
  \ind -- proof of one verification condition omitted
\end{numquote}\normalsize
Lines 4--7 remain unchanged and obtain the root $r$ of $x$ and the root $s$ of $y$.
If the two roots are identical, $x$ and $y$ are already in the same set and nothing needs to be done.
Otherwise either the parent of $r$ is set to $s$ in line 10 or the other way around in line 12, depending on how their ranks compare in line 9.
The node with smaller rank needs to point to the node with larger rank to obtain trees with smaller heights.
If the ranks of $r$ and $s$ are equal, as checked in line 13, the rank of the remaining root $r$ needs to be incremented, which happens in line 14.

\medskip
We show that the rank array satisfies three properties which are established or preserved by the union-find operations:
\begin{quote}
  rank\_property $p$ $\rank$ $\iff$ \\
  \ind mapping $\rank$ $\wedge$ $(p - \unit) \comp \rank \lleq \rank \comp \psuc^+$ $\wedge$ $|\kroots{p}| \leq |\cpl{\psuc^+ \comp \cnv{\rank} \comp \ltop}|$
\end{quote}
First, every node has a rank, that is, the rank array is a mapping.
This is non-trivial to maintain because only finitely many numbers can be represented.
Specifically, since the successor $\psuc$ is partial, the assignment in line 14 of union\_sets\_by\_rank would not increment the largest number that can be represented.
A proof that this rank property is maintained ensures that an overflow never occurs.

\medskip
Second, the rank of a node is strictly smaller than the rank of its parent, except if the node is a root.
This implies that the rank of a node is an upper bound on the height of its subtree.
The left-hand side $(p - \unit) \comp \rank$ relates a node to the rank of its parent if the parent is a different node.
The right-hand side $\rank \comp \psuc^+$ relates a node to all numbers greater than its rank.
The inclusion $(p - \unit) \comp \rank \lleq \rank \comp \psuc^+$ specifies that the parent's rank must be among those numbers.
We can verify this interpretation by deriving the logical meaning of this relation-algebraic formula:
{\small{\begin{align*}
  &      (p - \unit) \comp \rank \lleq \rank \comp \psuc^+ \\[-1.pt]
  & \iff \forall a, b : (a,b) \in (p - \unit) \comp \rank \implies (a,b) \in \rank \comp \psuc^+ \\[-1.pt]
  & \iff \forall a, b : (\exists c : (a,c) \in (p - \unit) \wedge (c,b) \in \rank) \implies (\exists d : (a,d) \in \rank \wedge (d,b) \in \psuc^+) \\[-1.pt]
  & \iff \forall a, b : (\exists c : (a,c) \in p \wedge (a,c) \notin \unit \wedge \rank(c) = b) \implies (\exists d : \rank(a) = d \wedge d < b) \\[-1.pt]
  & \iff \forall a, b : (\exists c : p(a) = c \wedge a \neq c \wedge \rank(c) = b) \implies \rank(a) < b \\[-1.pt]
  & \iff \forall a, b : a \neq p(a) \wedge \rank(p(a)) = b \implies \rank(a) < b \\[-1.pt]
  & \iff \forall a : a \neq p(a) \implies \rank(a) < \rank(p(a))
\end{align*} } } \normalsize \vspace*{-2mm}

Third, the number of roots in the disjoint-set forest (which is the number of disjoint sets) is not larger than $m-k$ where $m$ is the total number of nodes and $k$ is the maximum rank of any node.
This property is established since initially all ranks are $\opzero$ and every node is a root.
To show that this property is maintained, note that whenever the rank of a node increases by $1$ in line 14, the former root $s$ has been linked to $r$ in line 12 and therefore the number of roots has decreased by $1$.
In other cases the number of roots decreases without changing the maximum rank.
Because there is always at least one root, $1 \leq m-k$ ensures that the maximum rank will never exceed $m-1$, that is, ranks never overflow.

To compare the number of roots and $m-k$ we compare the cardinality of the set of roots and the set of numbers between $k$ inclusively and $m$ exclusively.
The following definition compares the cardinality of two sets using the existence of an injective univalent relation between the vectors $v$ and $w$ representing the sets:
\begin{quote}
  $|v| \leq |w| \iff \exists i$ . injective $i$ $\wedge$ univalent $i$ $\wedge$ $v \lleq i \comp w$
\end{quote}
See \cite{Kawahara2006} for a more comprehensive axiomatisation of the cardinalities of heterogeneous relations.
The vector $\psuc^+ \comp \cnv{\rank} \comp \ltop$ contains a number if there is a greater number that is the rank of a node.
These are just the numbers smaller than the maximum rank, so the complement of the vector gives the desired numbers between $k$ and $m$.
We can verify this interpretation by deriving the logical meaning of this relation-algebraic expression:
{\small{\begin{align*}
  &      (a,b) \in \cpl{\psuc^+ \comp \cnv{\rank} \comp \ltop} \\[-1.pt]
  & \iff \neg (a,b) \in \psuc^+ \comp \cnv{\rank} \comp \ltop \\[-1.pt]
  & \iff \neg \exists c, d : (a,c) \in \psuc^+ \wedge (c,d) \in \cnv{\rank} \wedge (d,b) \in \ltop \\[-1.pt]
  & \iff \neg \exists c, d : a < c \wedge (d,c) \in \rank \\[-1.pt]
  & \iff \neg \exists c, d : a < c \wedge \rank(d) = c \\[-1.pt]
  & \iff \neg \exists d : a < \rank(d) \\[-1.pt]
  & \iff \forall d : \rank(d) \leq a \\[-1.pt]
  & \iff k \leq a
\end{align*} }\normalsize }
\indent Proof that union-by-rank maintains the rank property involves showing that it is maintained by path compression, which changes the parent relation.
To this end we use the precise effect of path compression described in Section \ref{section.path-compression} and captured in its postcondition.
This avoids adding the rank property to the loop invariant of path compression.\vspace*{-2mm}

\section{Related work and limitations}
\label{section.related-work}

In this section we review work related to the contributions of this paper and discuss its limitations.

\subsection{Related work}\vspace*{-1mm}
\label{section.related}

The semantics of array or general state access is well understood and has been described in many different formalisms.
We discuss a selection of these related works.
An early example are the $a$ and $c$ functions in \cite{McCarthy1963} for updating and reading state vectors, which map variables to values.
Arrays are modelled as mappings in \cite{Hoare1972} and selective array updates are defined as updates of mappings.
Such updates are more formally defined in \cite{Reynolds1979}.
Functional and relational override in Z \cite{Spivey1989} dates back to \cite{Hayes1985}.
Overwriting one relation with another also appears in \cite{Moeller1993b} where it is used for pointer structures.
Axioms for state attributes and array access are given in \cite{BackWright1998}; some of these are used for lenses \cite{FosterGreenwaldMoorePierceSchmitt2007}.
A definition of general updates in Kleene algebras with domain is given in \cite{Ehm2004}.
The relation-algebraic semantics given in the present paper specialises definitions given in \cite{Spivey1989,Hayes1985,Moeller1993b,BackWright1998,FosterGreenwaldMoorePierceSchmitt2007,Ehm2004} to selective array updates studied in \cite{McCarthy1963,Hoare1972,Reynolds1979}.

Relation-algebraic methods have been used for the description and verification of numerous algorithms, in particular, on graphs as mentioned in the introduction.
Especially relevant to the present work on disjoint sets are relational formalisations of forests and reachability; for example, see \cite{Berghammer1999,Moeller1993,SchmidtStroehlein1989}.
Also relevant are relational models of stores modelling pointer structures and using relational overwrite operations \cite{Moeller1993,Moeller1993b,Moeller1997,Moeller1999}.

There are several formally verified implementations of disjoint-set forests.
A persistent version of the data structure is verified in Coq by \cite{ConchonFilliatre2007}.
The specification is in terms of predicate logic and the implementation is based on a mathematical model of ML including references.
See \cite{LammichMeis2012} for a verification using separation logic in Isabelle/HOL also based on a logical specification; the programs can be translated to executable code in various target languages.
Program complexity and correctness of an OCaml implementation is proved in \cite{ChargueraudPottier2019} using separation logic in Coq based on a predicate-logic specification.
See the latter paper for an overview of other formal verifications and further related works.
We also mention the manual derivation of implementations of union and find from an abstract specification by data refinement in \cite{Jones1980,Jones1990}; it does not integrate path compression or union-by-rank.
The present paper gives a relation-algebraic specification and proof, which does not cover complexity of the union and find operations.

\subsection{Limitations}
\label{section.limitations}

The approach used in this paper has two limitations.
First, we only cover arrays with the same index and value sets.
This is because Isabelle/HOL's class hierarchy currently only supports homogeneous
  relation algebras.
In this regard, we argue as follows.
Firstly, we value formally verified results which improve confidence compared to manual proofs.
Secondly, the choice of Isabelle/HOL is due to having extensive libraries for relation algebras and a high degree of automation.
Thirdly, the choice of using classes of algebraic structures instead of concrete relations follows our aim to further develop and strengthen algebraic techniques, which have been useful in other verification efforts.
Finally, this limitation is not fundamental but can be overcome by providing a library for heterogeneous relation algebras with automation support, however this is outside the scope of the current paper.

The second limitation is that we do not prove the time complexity of the disjoint-forest operations.
We are aware of two libraries for time-complexity proofs in Isabelle/HOL: Imperative HOL with Time \cite{ZhanHaslbeck2018} and LLVM with Time \cite{HaslbeckLammich2022}.
Currently, neither is part of the Archive of Formal Proofs or sufficiently documented to be easily applicable to new projects.
Again, this is not a fundamental limitation.

\section{Conclusion}

This paper has given a simple relation-algebraic semantics for read and write operations on associative arrays.
Based on this semantics, we added such operations to a sequential programming language used for specifying and verifying programs in Isabelle/HOL.
We implemented disjoint-set forests with path compression, path splitting, path halving and union-by-rank this way and proved their correctness.

\medskip
We have used single relational updates to express the precise effect of various path-compression techniques.
This was helpful for specification and for reasoning about the lower-level implementations.
It could also feature in refinement-based program development \cite{ArmstrongGomesStruth2016}.

To deal with union-by-rank it was necessary to extend relation algebras by basic arithmetic.
This is usually done with heterogeneous relations which are, however, not supported by the current relation-algebraic framework in Isabelle/HOL.
On the one hand, this required working in an alternative model that represents only a restricted subset of numbers.
On the other, such a setting shares properties with fixed-size number types available in many programming languages for which it can be important to show that overflows do not occur.

Examples in Section \ref{section.motivation} show that Kleene relation algebras can be used to express complex logical specifications in compact, readable and understandable ways.
Another such example is the final condition $\awrite{p_0}{(\anc{p_0}{x} - \anc{p_0}{w})}{y} = p$ of the path compression invariant in Section \ref{section.path-compression}, which expresses the effect of path compression using a single relational update.
Further examples are the conditions describing the effect of path splitting and path halving using relational updates in Sections \ref{section.path-splitting} and \ref{section.path-halving}.
The rank properties discussed in Section \ref{section.union-by-rank} require more familiarity with expressing specifications using relational operations.
In these cases it is always possible to link relation-algebraic expressions to their logical or set-theoretic interpretation, which helps constructing and understanding such formulas.

In this paper we have focused on data represented by associative arrays.
Relation algebras can be extended by other types of data using the domain constructions of products, sums, quotients, subsets and power sets discussed, for example, in \cite{Schmidt2011}.
Various data types and their application to sorting algorithms, optimisation problems, dynamic programming and greedy algorithms in a relation-algebraic framework are studied in \cite{BirdMoor1997}.

A future task is to integrate the implementation given in this paper with relation-algebraic implementations of Kruskal's minimum spanning tree algorithm.
For this reason our Isabelle/HOL theory uses Stone-Kleene relation algebras, which are weaker than Kleene relation algebras and can represent weighted graphs \cite{Guttmann2018c}.
While Kruskal's algorithm has already been formally verified in Isabelle/HOL \cite{HaslbeckLammichBiendarra2019}, the relation-algebraic specification and implementation mean that the same algorithm is correct for various optimisation problems as discussed in Section \ref{section.motivation}.
A further direction of research is to consider how relation-algebraic methods can support complexity analysis of algorithms.

\subsection*{Acknowledgement}

I thank the anonymous referees for their helpful feedback.

\appendix

\section{Further properties of forests}
\label{section.forest-properties}

In this appendix we discuss results used for verifying the correctness of path compression, path splitting, path halving and union-sets.

We first discuss a selection of results used for maintaining the invariant for path compression in Section \ref{section.path-compression}.
Part of the maintenance is to show that the parent relation (without self-loops) remains acyclic.
Path compression updates the parent relation by letting the parents of visited nodes point to the root of the tree.
Part 1 of the following theorem shows that updating the parent of a node $w$ to any ancestor $y$ of $w$ does not introduce cycles (ignoring self-loops).

\begin{thm}
  \label{theorem.path-compression}
  \begin{enumerate}
  \item[]
  \item $\awrite{p}{w}{y} - \unit$ is acyclic if $p - \unit$ is acyclic, $w$ and $y$ are points and $y \lleq \anc{p}{w}$.\afplemma{update_acyclic_1}
        \label{theorem.path-compression.update-1}
  \item $x \meet p^* = (x \meet \unit) \join ((x \meet p) \comp (\cpl{x} \meet p)^*)$ if $x$ is a point.\afplemma{omit_redundant_points}
        \label{theorem.path-compression.optimise-1}
  \item $x \meet p^* = (x \meet \unit) \join (x \meet (p \meet \cnv{\cpl{x}})^+)$ if $x$ is a point.\afplemma{omit_redundant_points_3}
        \label{theorem.path-compression.optimise-2}
  \item $x \meet y = \lbot$ if $x$ and $y$ are points such that $x \neq y$.\afplemma{distinct_points}
        \label{theorem.path-compression.distinct-points}
  \end{enumerate}
\end{thm}

\begin{prf}
  See Appendix \ref{section.proofs} for proofs of Theorems \ref{theorem.path-compression}.\ref{theorem.path-compression.update-1} and \ref{theorem.path-compression}.\ref{theorem.path-compression.distinct-points}, and \cite{Guttmann2020c} for all proofs.
\end{prf}

Parts 2 and 3 optimise iterations similar to \cite[Lemma 4]{Moeller1993}; for related techniques see also \cite{BackhouseCarre1975}.
The element $x \meet p^*$ on the left-hand side relates the node $x$ to all nodes reachable from it by zero or more steps in the graph $p$.
The right-hand side of part 2 contains $x \meet \unit$, which relates $x$ to itself, and $(x \meet p) \comp (\cpl{x} \meet p)^*$, which relates $x$ to nodes reachable from it by one step in $p$ followed by zero or more steps in $\cpl{x} \meet p$.
This means that edges starting in $x$ have to be considered at most in the first step and can be omitted in the remaining steps.
In maintaining the invariant, this is applied with $x = w$, so that the remaining steps only use edges not starting in $w$, which is important since these edges are not affected by the update to the forest.

Part 4 of the previous theorem ultimately derives from the Tarski rule and states that different points are disjoint as relations.
This is a general result used in several arguments; we explain one of them.
In maintaining the invariant, we need to show that updating $p$ does not change the set of its roots.
The update changes $p$ at index $w$ to the new value $y$, so this part of $p$ changes from $w \meet p$ to $w \meet \cnv{y}$.
The roots in this part are $w \meet p \meet \unit$ and $w \meet \cnv{y} \meet \unit$ and we show that both expressions are $\lbot$.
To this end, observe that $y \neq w$ since $y$ is a root according to the precondition, but the parent of $w$ is different from $y$ according to the condition of the while-loop.
First, $w \meet p \meet \unit \lleq w \meet \unit = \lbot$ because the node $w$ does not have a self-loop; otherwise $y = w$ would hold since $y$ is reachable from $w$ according to the loop invariant.
Second, $w \meet \cnv{y} \meet \unit = w \meet y \meet \unit \lleq w \meet y = \lbot$ by a general property of relation algebras and part 4 of the previous theorem.

\medskip
Next, we present properties involving the grandparent relation, which are used for verifying the correctness of path splitting and path halving in Sections \ref{section.path-splitting} and \ref{section.path-halving}.

\begin{thm}
  \label{theorem.square}
  \begin{enumerate}
  \item[]
  \item $\cnv{(x \comp x)}\astop \meet (\cnv{x} \comp \cnv{(x \comp x)}\astop) = (\unit \meet x) \comp (\cnv{(x \comp x)}\astop \meet (\cnv{x} \comp \cnv{(x \comp x)}\astop))$ if $x$ is univalent and $x - \unit$ is acyclic.\afplemma{even_odd_root}
  \item $x^+ \meet \unit = (x \comp x) \meet \unit = x \meet \unit$ if $x - \unit$ is acyclic.\afplemma{acyclic_square}
  \item $(x \comp x) - \unit \lleq x^* - \unit \lleq (x - \unit)^+$.\afplemma{star_irreflexive_part}
        \label{theorem.square.square-irreflexive-part}
  \item $\awrite{x}{y}{\aread{x}{\aread{x}{y}}} \lleq x \join (x \comp x)$ if $y$ is a point.\afplemma{update_square}
  \item $\awrite{x}{y}{\aread{x}{\aread{x}{y}}} \meet \unit = x \meet \unit$ if $x - \unit$ is acyclic and $y$ is a point.\afplemma{diagonal_update_square}
  \item $\fc(\awrite{x}{y}{\aread{x}{\aread{x}{y}}}) = \fc(x)$ if $x$ is a mapping and $y$ is a point.\afplemma{fc_update_square}
  \item $\awrite{x}{y}{\cnv{(x \comp x)}} \lleq x \join (x \comp x)$.\afplemma{update_square_ub}
        \label{theorem.square.update-square-ub}
  \item $\awrite{x}{y}{\cnv{(x \comp x)}} - \unit$ is acyclic if $x - \unit$ is acyclic.\afplemma{acyclic_update_square}
        \label{theorem.square.acyclic-update-square}
  \item $\awrite{x}{y}{\cnv{(x \comp x)}}$ is a forest if $x$ is a forest and $y$ is a vector.\afplemma{disjoint_set_forest_update_square}
        \label{theorem.square.disjoint-set-forest-update-square}
  \end{enumerate}
\end{thm}

\begin{prf}
  See Appendix \ref{section.proofs} for proofs of Theorems \ref{theorem.square}.\ref{theorem.square.square-irreflexive-part}, \ref{theorem.square}.\ref{theorem.square.update-square-ub}, \ref{theorem.square}.\ref{theorem.square.acyclic-update-square} and \ref{theorem.square}.\ref{theorem.square.disjoint-set-forest-update-square}, and \cite{Guttmann2020c} for all proofs.
\end{prf}

Part 1 shows that if between two nodes of a forest there is a path of even length and one of odd length, the start node must be a root.
Part 2 shows that the self-loops of the grandparent relation are the self-loops of the parent relation.
Parts 5--6 show that updating nodes to their grandparents does not change the semantics of the disjoint-set forest or its roots.
Parts 8--9 show that such updates maintain the forest property.

\medskip
Finally, we discuss a selection of results used for proving the correctness of union-sets in Section \ref{section.union-sets}.
Part 1 of the following result is similar to \cite[Proposition 3]{BerghammerStruth2010}; also see \cite[Section 4.3]{BackhouseCarre1975}.
It considers reachability under the union of two relations $x$ and $y$, where $x$ is an arc containing just one edge.
It then suffices to use the edge $x$ at most once: $y^+$ describes the case where $x$ is not needed and $y^* \comp x \comp y^*$ describes the case where $x$ is used once, preceded and followed by any number of \linebreak edges in $y$.

\begin{thm}
  \label{theorem.union-sets}
  \begin{enumerate}
  \item[]
  \item $(x \join y)^+ = y^+ \join (y^* \comp x \comp y^*)$ if $x$ is an arc.\afplemma{plus_arc_decompose}
        \label{theorem.union-sets.plus-arc-decompose}
  \item $\awrite{p}{w}{y} - \unit$ is acyclic if $p - \unit$ is acyclic, $w$ and $y$ are points, and $y \meet (p^* \comp w) = \lbot$.\afplemma{update_acyclic_4}
        \label{theorem.union-sets.update-acyclic-4}
  \item $\awrite{p}{w}{w} - \unit$ is acyclic if $p - \unit$ is acyclic and $w$ is a point.\afplemma{update_acyclic_5}
  \end{enumerate}
\end{thm}

\begin{prf}
  See Appendix \ref{section.proofs} for proofs of Theorems \ref{theorem.union-sets}.\ref{theorem.union-sets.plus-arc-decompose} and \ref{theorem.union-sets}.\ref{theorem.union-sets.update-acyclic-4}, and \cite{Guttmann2020c} for all proofs.
\end{prf}

Parts 2 and 3 are similar to Theorem \ref{theorem.path-compression}.\ref{theorem.path-compression.update-1}.
In part 2 the parent of $w$ is updated to a node $y$ from which $w$ is not reachable in $p$.
This does not introduce a cycle (ignoring self-loops).
Part 3 shows that creating a self-loop on $w$ does not introduce a cycle (ignoring self-loops).

These results are used to show that the assignment in line 8 of union-sets maintains the forest property.
If the arguments $x$ and $y$ of union-sets are in the same tree, the roots $r$ and $s$ will be equal, so line 8 creates a self-loop and the correctness proof uses part 3 of the preceding result.
Alternatively, it could be proved that the assignment in line 8 does not change the forest in this case.
Part 2 of the preceding result is used if nodes $x$ and $y$ are in different trees.

\section{Example proofs}
\label{section.proofs}

All results of this paper are proved in the Isabelle/HOL theory files available in the Archive of Formal Proofs \cite{Guttmann2020c}.
In this appendix we prove some results as examples.

\medskip
We first list a number of laws of Kleene relation algebras used in the proofs below.
They are grouped by their assumptions and labelled for reference.
In this list, $u, v, x, y, z \in S$ for a Kleene relation algebra $S$.
\begin{enumerate}
\renewcommand\labelenumi{\theenumi}
\renewcommand{\theenumi}{(\arabic{enumi})}
\item Let $u$ and $v$ be vectors.
      Then (1.1) $u \meet (x \comp y) = (u \meet x) \comp y$ and (1.2) $(x \meet \cnv{u}) \comp y = x \comp (u \meet y)$ and (1.3) $x \comp (y \meet \cnv{u}) = (x \comp y) \meet \cnv{u}$.
      Moreover (1.4) $u \meet v$ and $x \comp u$ and $\cpl{u}$ are vectors and (1.5) $u \meet \cnv{v} = u \comp \cnv{v}$.
\item Let $u \lleq \unit$.
      Then (2.1) $(u \comp \ltop) \meet x = u \comp x$ and $\cnv{u} = u$, hence (2.2) $x \meet \unit = \cnv{x} \meet \unit$.
\item Let $u \comp v \lleq \unit$.
      Then $(u \join v)^* = v^* \comp u^*$.
\item Let $u \lleq v$ and let $v$ be injective.
      Then $u = v \meet (\ltop \comp u)$.
\item $x$ is surjective if and only if $\ltop \comp x = x$.
\item Let $u$ and $v$ be bijective.
      Then (6.1) $u \lleq x \comp v \iff v \lleq \cnv{x} \comp u$ and (6.2) $x \lleq y \comp u \iff x \comp \cnv{u} \lleq y$ and (6.3) $x \lleq \cnv{u} \comp y \iff u \comp x \lleq y$.
\item Let $u$ be univalent.
      Then $u \comp (x \meet y) = (u \comp x) \meet (u \comp y)$.
\item Let $u$ and $v$ be mappings.
      Then (8.1) $u \comp v$ is a mapping and (8.2) $\cpl{u \comp x} = u \comp \cpl{x}$.
\item Let $u$ be a mapping and let $v$ be a point.
      Then $u \meet v$ is an arc.
\item Let $u$ be an arc.
      Then $u \comp x \comp u \lleq u$.
\item (11.1) $x \meet y \lleq z \iff x \lleq z \join \cpl{y}$ and (11.2) $x \comp y \lleq z \iff \cnv{x} \comp \cpl{z} \lleq \cpl{y}$ and (11.3) $x \comp y \lleq \cpl{\unit} \iff y \comp x \lleq \cpl{\unit}$.
\item (12.1) $x^* \comp x^* = x^*$ and (12.2) $x^*\astop = x^*$ and (12.3) $x^+\plusop = x^+$.
\end{enumerate}
We furthermore use properties of semirings, lattices, Boolean algebras and partial orders without explicitly mentioning them.
Details are available in the Isabelle/HOL proofs.

Results are grouped by content not by the order in which they are proved.
Some proofs refer to later results; formal verification of course ensures that there are no cyclic dependencies.

\begin{prfof}[of Theorem \ref{theorem.awrite-aread}.\ref{theorem.awrite-aread.aread-iff}]
  For the backward implication assume $y \meet x = y \comp \cnv{z}$, then
  \begin{align*}
    \aread{x}{y} = \cnv{x} \comp y = (\cnv{x} \meet \cnv{y}) \comp \ltop = \cnv{(x \meet y)} \comp \ltop = \cnv{(y \comp \cnv{z})} \comp \ltop = z \comp \cnv{y} \comp \ltop = z \comp \ltop \comp y = z \comp \ltop = z
  \end{align*}
  using the definition of array read, (1.2), distributivity of transposition over meet, the assumption, distributivity of transposition over composition and involution of transposition, (1.2), (5) and that $z$ is a vector.

\medskip
  For the forward implication assume $\cnv{x} \comp y = z$, then we show $y \meet x \lleq y \comp \cnv{z}$ and $y \comp \cnv{z} \lleq y \meet x$.
  First,
  \begin{align*}
    y \meet x \lleq y \meet (\ltop \comp (y \meet x)) = y \meet (\cnv{y} \comp x) = y \meet \cnv{(\cnv{x} \comp y)} = y \meet \cnv{z} = y \comp \cnv{z}
  \end{align*}
  using that composition with $\ltop$ is $\lleq$-increasing, (1.2), distributivity of transposition over composition and involution of transposition, the assumption and (1.5).
  Second,
  \begin{align*}
    y \comp \cnv{z} = y \comp \cnv{(\cnv{x} \comp y)} = y \comp \cnv{y} \comp x \lleq y \comp \ltop \meet x = y \meet x
  \end{align*}
  using the assumption, distributivity of transposition over composition and involution of transposition, injectivity of $y$ and that $y$ is a vector.

  In Isabelle/HOL, the user's contribution is to reduce the overall goal to the three subgoals discussed above, which is a natural way of breaking the proof into smaller chunks.
  Then each of the three subgoals is proved automatically by Sledgehammer using integrated automated theorem provers or SMT solvers \cite{PaulsonBlanchette2010}.
\end{prfof}

\begin{prfofnoqed}[of Theorem \ref{theorem.awrite-aread-2}.\ref{theorem.awrite-aread-2.update-injective-swap}]
  By definition of array write and read,
  \begin{align*}
    \awrite{x}{y}{\aread{x}{z}} = \awrite{x}{y}{\cnv{x} \comp z} = (y \meet \cnv{(\cnv{x} \comp z)}) \join (\cpl{y} \meet x) = (y \meet (\cnv{z} \comp x)) \join (\cpl{y} \meet x)
  \end{align*}
  and therefore
  \begin{align*}
    \awrite{\awrite{x}{y}{\aread{x}{z}}}{z}{\aread{x}{y}} = (z \meet (\cnv{y} \comp x)) \join (\cpl{z} \meet y \meet (\cnv{z} \comp x)) \join (\cpl{z} \meet \cpl{y} \meet x) = u_1 \join u_2 \join u_3
  \end{align*}
  for $u_1 = z \meet (\cnv{y} \comp x)$ and $u_2 = \cpl{z} \meet y \meet (\cnv{z} \comp x)$ and $u_3 = \cpl{z} \meet \cpl{y} \meet x$.
  To show that $u_1 \join u_2 \join u_3$ is injective, it thus remains to show $u_i \comp \cnv{u_j} \lleq \unit$ for each $i, j \in \{ 1, 2, 3 \}$:
  \begin{enumerate}
  \item $u_1 \comp \cnv{u_1} \lleq z \comp \cnv{z} \lleq \unit$ since $z$ is injective.
  \item $u_1 \comp \cnv{u_2} \lleq u_1 \comp (\cnv{y} \meet (\cnv{x} \comp z)) = (z \meet (\cnv{z} \comp x)) \comp (\cnv{y} \meet (\cnv{x} \comp y)) = (z \meet \cnv{z}) \comp x \comp \cnv{x} \comp (\cnv{y} \meet y) \lleq \unit$ using (1.4) and (1.2), (1.1) and (1.3), (1.5) and injectivity of $x$, $y$ and $z$.
  \item $u_1 \comp \cnv{u_3} \lleq \cnv{y} \comp x \comp (\cnv{\cpl{y}} \meet \cnv{x}) = (\cnv{y} \comp x \comp \cnv{x}) \meet \cnv{\cpl{y}} \lleq \cnv{y} \meet \cnv{\cpl{y}} = \lbot \lleq \unit$ using (1.4) and (1.3), and injectivity of $x$.
  \item $u_2 \comp \cnv{u_1} \lleq \unit$ follows by transposition from subgoal 2.
  \item $u_2 \comp \cnv{u_2} \lleq y \comp \cnv{y} \lleq \unit$ since $y$ is injective.
  \item $u_2 \comp \cnv{u_3} \lleq \cnv{z} \comp x \comp (\cnv{\cpl{z}} \meet \cnv{x}) = (\cnv{z} \comp x \comp \cnv{x}) \meet \cnv{\cpl{z}} \lleq \cnv{z} \meet \cnv{\cpl{z}} = \lbot \lleq \unit$ similarly to the proof of subgoal 3.
  \item $u_3 \comp \cnv{u_1} \lleq \unit$ follows by transposition from subgoal 3.
  \item $u_3 \comp \cnv{u_2} \lleq \unit$ follows by transposition from subgoal 6.
  \item $u_3 \comp \cnv{u_3} \lleq x \comp \cnv{x} \lleq \unit$ since $x$ is injective.
  \QED
  \end{enumerate}
\end{prfofnoqed}

\begin{prfof}[of Theorem \ref{theorem.wcc}.\ref{theorem.wcc.closure}]
  First, $\wcc(x)$ is symmetric by $(x \join \cnv{x})^*\cnvop = \cnv{(x \join \cnv{x})}\astop = (\cnv{x} \join \cnv{\cnv{x}})^* = (x \join \cnv{x})^*$ using that Kleene star and transposition commute.
  Hence $(\wcc(x) \join \cnv{\wcc(x)})^* = \wcc(x)^* = (x \join \cnv{x})^*\astop = (x \join \cnv{x})^*$ using (12.2), which shows that $\wcc$ is idempotent.
  Next, $x \lleq x \join \cnv{x} \lleq (x \join \cnv{x})^*$ shows that $\wcc$ is $\lleq$-increasing.
  Finally, $\wcc$ is $\lleq$-isotone since $\join$, $\cnvop$ and $\astop$ are $\lleq$-isotone.
\end{prfof}

\begin{prfof}[of Theorem \ref{theorem.component}.\ref{theorem.component.equivalence}]
  Reflexivity of $\fc(x)$ follows by $\unit = \unit \comp \unit \lleq x^* \comp \cnv{x}\astop$ using reflexivity of the Kleene star.
  Transitivity of $\fc(x)$ follows by $x^* \comp \cnv{x}\astop \comp x^* \comp \cnv{x}\astop \lleq (x \join \cnv{x})^* = x^* \comp \cnv{x}\astop$ using (12.1) and (3) since $x$ is univalent.
  Symmetry of $\fc(x)$ follows by $\cnv{(x^* \comp \cnv{x}\astop)} = \cnv{\cnv{x}\astop} \comp \cnv{x^*} = \cnv{\cnv{x}}\astop \comp \cnv{x}\astop = x^* \comp \cnv{x}\astop$ using that Kleene star and transposition commute.
\end{prfof}

\begin{prfof}[of total correctness of find-set]
  The invariant comprises three conditions.
  The first two are established immediately from the precondition.
  The remaining condition $x \lleq \anc{p}{x} = \cnv{p}\astop \comp x$ follows by reflexivity of the Kleene star.

\medskip
  For maintaining the invariant we need to show that it holds for $\aread{p}{y}$ assuming that it holds for $y$ and $y \neq \aread{p}{y}$ indicating the while-loop has not yet terminated.
  The first condition for $\aread{p}{y}$ follows immediately from the assumption for $y$.
  The second condition requires that $\aread{p}{y}$ is a point, which follows by Theorem \ref{theorem.awrite-aread}.\ref{theorem.awrite-aread.aread-point} since $y$ is a point and $p$ is a forest.
  The third condition follows by
  \begin{align*}
    \aread{p}{y} = \cnv{p} \comp y \lleq \cnv{p} \comp \cnv{p}\astop \comp x \lleq \cnv{p}\astop \comp x = \anc{p}{x}
  \end{align*}
  using $y \lleq \anc{p}{x}$ and properties of the Kleene star.

\medskip
  For total correctness we prove $\{ z \mid z \lleq \anc{p}{\aread{p}{y}} \} \subset \{ z \mid z \lleq \anc{p}{y} \}$, which implies that the variant decreases since these sets are finite.
  The non-strict inclusion follows from
  \begin{align*}
    \anc{p}{\aread{p}{y}} = \cnv{p}\astop \comp \cnv{p} \comp y \lleq \cnv{p}\astop \comp y = \anc{p}{y}
  \end{align*}
  For strict inclusion we show $y$ is in the difference of the two sets.
  First, $y$ is in the bigger set since $y \lleq \cnv{p}\astop \comp y = \anc{p}{y}$ by reflexivity of the Kleene star.
  Second, we derive a contradiction from assuming that $y$ is in the smaller set.
  Since $y$ and $\aread{p}{y}$ are different points, we have $(\cnv{p} \comp y) \meet y = \aread{p}{y} \meet y = \lbot$ by Theorem \ref{theorem.path-compression}.\ref{theorem.path-compression.distinct-points}, and therefore
  \begin{align*}
    \cnv{p} \comp y = ((\cnv{p} \meet \unit) \comp y) \join ((\cnv{p} - \unit) \comp y) \lleq ((\cnv{p} \comp y) \meet y) \join ((\cnv{p} - \unit) \comp y) = (\cnv{p} - \unit) \comp y
  \end{align*}
  So $y$ being in the smaller set yields
  \begin{align*}
    y \lleq \anc{p}{\aread{p}{y}} = \cnv{p}\astop \comp \cnv{p} \comp y = (\cnv{p} - \unit)^* \comp \cnv{p} \comp y \lleq (\cnv{p} - \unit)^+ \comp y
  \end{align*}
  Hence $y \comp \cnv{y} \lleq (\cnv{p} - \unit)^+ \lleq \cpl{\unit}$ using (6.2) since $y$ is a point and that $p$ is a forest.
  With injectivity of $y$ this implies $y = \lbot$, which contradicts that $y$ is a point.

\medskip
  Finally we show that the invariant implies the postcondition when the while-loop terminates at $y = \aread{p}{y}$.
  The first postcondition, $y$ is a point, is immediate from the invariant.
  It remains to show $y = \kroot{p}{x}$.
  Observe that $y = \aread{p}{y} = \cnv{p} \comp y$ implies $y \comp \cnv{y} \lleq \cnv{p}$ using (6.2), whence $y \comp \cnv{y} \lleq p \meet \unit$ by transposition and since $y$ is injective.
  Hence $y \lleq (p \meet \unit) \comp y \lleq \kroots{p}$ using (6.2), which together with $y \lleq \anc{p}{x}$ from the invariant establishes $y \lleq \kroot{p}{x}$.
  For the opposite inequality, first note $x \lleq p^* \comp y$ is a consequence of the invariant using (6.1).
  Second,
  \begin{align*}
    \cnv{p} \comp p^* \comp y = (\cnv{p} \comp p \comp p^* \comp y) \join (\cnv{p} \comp y) \lleq (p^* \comp y) \join y = p^* \comp y
  \end{align*}
  since $p$ is univalent.
  Together we obtain $\cnv{p}\astop \comp x \lleq p^* \comp y$ by Kleene star induction.
  We also have $(p \comp y) \meet \kroots{p} = (p \meet \unit) \comp p \comp y \lleq \cnv{p} \comp p \comp y \lleq y$ using (2.1) and (2.2) and that $p$ is univalent.
  Therefore $p \comp y \lleq y \join \cpl{\kroots{p}}$ using (11.1).
  Next, $p \comp \cpl{\kroots{p}} \lleq \cpl{\kroots{p}}$ by Theorem \ref{theorem.root}.\ref{theorem.root.roots-successor-loop} and (11.2).
  Hence $p \comp (y \join \cpl{\kroots{p}}) = (p \comp y) \join (p \comp \cpl{\kroots{p}}) \lleq y \join \cpl{\kroots{p}}$.
  Kleene star induction and a previous fact imply $\cnv{p}\astop \comp x \lleq  p^* \comp y \lleq y \join \cpl{\kroots{p}}$.
  Using (11.1) we conclude $\kroot{p}{x} \lleq y$.
\end{prfof}

\begin{prfof}[of Theorems \ref{theorem.z-s}.\ref{theorem.z-s.s-inf-1-below-z} and \ref{theorem.z-s}.\ref{theorem.z-s.top-s-sup-conv-z}]
  For Theorem \ref{theorem.z-s}.\ref{theorem.z-s.s-inf-1-below-z} we have $(\suc \meet \unit) \comp \cnv{\suc} \lleq (\suc \comp \cnv{\suc}) \meet \cnv{\suc} \lleq \unit \meet \cnv{\suc} = \suc \meet \unit$ using injectivity of $\suc$ and (2.2).
  Hence $(\suc \meet \unit) \comp \cnv{\suc}\astop \lleq \suc \meet \unit$ by Kleene star induction.
  Thus
  \begin{align*}
    \suc \meet \unit \lleq (\suc \meet \unit) \comp \ltop = (\suc \meet \unit) \comp \cnv{\suc}\astop \comp \zero \lleq (\suc \meet \unit) \comp \zero \lleq \unit \comp \zero = \zero
  \end{align*}
  using Peano axiom $\cnv{\suc}\astop \comp \zero = \ltop$.
  Theorem \ref{theorem.z-s}.\ref{theorem.z-s.top-s-sup-conv-z} immediately follows from this Peano axiom by unfolding the Kleene star.
\end{prfof}

\begin{prfof}[of Theorem \ref{theorem.s'}.\ref{theorem.s'.s'-star-z-top}]
  With $\maxi = \suc \comp \zero$ we have
  \begin{align*}
    \ltop = \cnv{\suc}\astop \comp \zero = \cnv{((\suc - \maxi) \join (\suc \meet \maxi))}\astop \comp \zero = \cnv{(\psuc \join (\suc \meet \maxi))}\astop \comp \zero
  \end{align*}
  using Theorem \ref{theorem.m-z}.\ref{theorem.m-z.s'-var}.
  If $\maxi = \lbot$, then $\cnv{(\psuc \join (\suc \meet \maxi))}\astop \comp \zero = \cnv{\psuc}\astop \comp \zero$ and we are done.
  Otherwise $\maxi$ is a point by Theorem \ref{theorem.m}.\ref{theorem.m.m-point-or-bot} and hence $\suc \meet \maxi$ is an arc using (9).
  Moreover $\suc - \cnv{\zero} = \psuc = \suc - \maxi \lleq \cpl{\maxi}$ by Theorem \ref{theorem.m-z}.\ref{theorem.m-z.s'-var} and therefore $\suc \meet \maxi \lleq \cnv{\zero}$ using (11.1).
  Hence
  \begin{align*}
    \cnv{(\psuc \join (\suc \meet \maxi))}\astop \comp \zero & = (\cnv{\psuc}\astop \comp \cnv{(\suc \meet \maxi)} \comp \cnv{\psuc}\astop \comp \zero) \join (\cnv{\psuc}\astop \comp \zero) \lleq (\cnv{\psuc}\astop \comp \zero \comp \ltop) \join (\cnv{\psuc}\astop \comp \zero) \\
    & = \cnv{\psuc}\astop \comp \zero
  \end{align*}
  using \cite[Proposition 3]{BerghammerStruth2010} (see proof of Theorem \ref{theorem.union-sets}.\ref{theorem.union-sets.plus-arc-decompose} in this appendix) and that $\zero$ is a vector.
\end{prfof}

\begin{prfof}[of Theorem \ref{theorem.m}]
  For the forward implication of Theorem \ref{theorem.m}.\ref{theorem.m.m-point-iff-s-surjective} assume $\maxi$ is a point.
  Hence $\unit \lleq \cnv{\zero} \comp \cnv{\suc} \comp \suc \comp \zero$, which implies $\zero \lleq \cnv{\suc} \comp \suc \comp \zero \lleq \cnv{\suc} \comp \ltop$ using (6.3).
  It follows that $(\cnv{\suc} \comp \cnv{\suc} \comp \ltop) \join \zero \lleq \cnv{\suc} \comp \ltop$.
  Therefore $\ltop = \cnv{\suc}\astop \comp \zero \lleq \cnv{\suc} \comp \ltop$ by a Peano axiom and Kleene star induction.
  Thus $\suc$ is surjective using (5).

  For the backward implication assume $\suc$ is surjective.
  Hence $\maxi = \suc \comp \zero$ is a point: it is bijective using (8.1) since both $\suc$ and $\zero$ are bijective, and it is a vector using (1.4) since $\zero$ is a vector.

  For the forward implication of Theorem \ref{theorem.m}.\ref{theorem.m.m-bot-iff-s-not-surjective} assume that $\maxi \neq \lbot$.
  Hence $\ltop \comp \suc \comp \zero = \ltop$ using the Tarski rule since $\zero$ is a vector.
  Therefore $\cnv{\zero} \lleq \ltop \comp \suc$ using (6.2) since $\zero$ is a point.
  Thus $\ltop = (\ltop \comp \suc) \join \cnv{\zero} = \ltop \comp \suc$ using Theorem \ref{theorem.z-s}.\ref{theorem.z-s.top-s-sup-conv-z}, whence $\suc$ is surjective using (5).

  For the backward implication assume $\suc$ is surjective.
  Hence $\maxi$ is a point by part 1 and therefore $\maxi \neq \lbot$.

  Theorem \ref{theorem.m}.\ref{theorem.m.m-point-or-bot} follows immediately from parts 1 and 2.
\end{prfof}

\begin{prfof}[of Theorem \ref{theorem.m-z}.\ref{theorem.m-z.s'-var}]
  We obtain
  \begin{align*}
    \psuc = \suc - \cnv{\zero} = \suc \comp (\unit - \cnv{\zero}) = \suc \comp (\unit - \zero) = (\suc \comp \unit) \meet (\suc \comp \cpl{\zero}) = \suc \meet \cpl{\suc \comp \zero} = \suc - \maxi
  \end{align*}
  using (1.4) and (1.3), (2.2), (7) and (8.2).
\end{prfof}

\begin{prfof}[of Theorems \ref{theorem.path-compression}.\ref{theorem.path-compression.update-1} and \ref{theorem.path-compression}.\ref{theorem.path-compression.distinct-points}]
  For Theorem \ref{theorem.path-compression}.\ref{theorem.path-compression.update-1}, since $w$ and $y$ are points, $y \lleq \anc{p}{w}$ is equivalent to $w \meet \cnv{y} = w \comp \cnv{y} \lleq p^*$ using (1.5), (6.1) and (6.2).
  Hence
  \begin{align*}
    w \meet \cnv{y} - \unit \lleq p^* - \unit = (p - \unit)^* - \unit = (p - \unit)^+ - \unit \lleq (p - \unit)^+
  \end{align*}
  Thus
  \begin{align*}
    \awrite{p}{w}{y} - \unit = (w \meet \cnv{y} - \unit) \join (\cpl{w} \meet p - \unit) \lleq (p - \unit)^+ \join (p - \unit) = (p - \unit)^+
  \end{align*}
  Therefore $(\awrite{p}{w}{y} - \unit)^+ \lleq (p - \unit)^+\plusop = (p - \unit)^+ \lleq \cpl{\unit}$ using (12.3).

  For Theorem \ref{theorem.path-compression}.\ref{theorem.path-compression.distinct-points}, we first show that $x \lleq y$ or $x \lleq \cpl{y}$ for any point $x$ and any vector $y$.
  Assume $x \lleq \cpl{y}$ does not hold, then $x \meet y \neq \lbot$ using (11.1), hence $\ltop \comp (x \meet y) = \ltop \comp (x \meet y) \comp \ltop = \ltop$ using (1.4) and the Tarski rule.
  Hence $x \meet y = x \meet (\ltop \comp (x \meet y)) = x \meet \ltop = x$ using (4) since $x \meet y \lleq x$ and $x$ is injective.
  Thus $x \lleq y$.

  We apply this both ways to points $x$ and $y$.
  If $x \neq y$, we cannot get both $x \lleq y$ and $y \lleq x$, so $x \lleq \cpl{y}$ or $y \lleq \cpl{x}$ must hold.
  Either is equivalent to $x \meet y = \lbot$ using (11.1).
\end{prfof}

\begin{prfof}[of Theorems \ref{theorem.square}.\ref{theorem.square.square-irreflexive-part}, \ref{theorem.square}.\ref{theorem.square.update-square-ub}, \ref{theorem.square}.\ref{theorem.square.acyclic-update-square} and \ref{theorem.square}.\ref{theorem.square.disjoint-set-forest-update-square}]
  For Theorem \ref{theorem.square}.\ref{theorem.square.square-irreflexive-part} observe that
  \begin{align*}
    x \comp x & = ((x \meet \unit) \comp x) \join ((x - \unit) \comp x) \lleq x \join ((x - \unit) \comp x) \lleq \unit \join (x - \unit) \join ((x - \unit) \comp x) \\
    & = ((x - \unit) \comp (x \join \unit)) \join \unit
  \end{align*}
  which implies
  \begin{align*}
    (x \comp x) - \unit \lleq (x - \unit) \comp (x \join \unit) = ((x - \unit) \comp (x - \unit)) \join (x - \unit) \lleq (x - \unit)^+
  \end{align*}
  using (11.1).
  Theorem \ref{theorem.square}.\ref{theorem.square.update-square-ub} follows from Theorem \ref{theorem.awrite-aread-2}.\ref{theorem.awrite-aread-2.update-ub} since transposition is involutive.
  As a consequence we obtain Theorem \ref{theorem.square}.\ref{theorem.square.acyclic-update-square} since
  \begin{align*}
    (\awrite{x}{y}{\cnv{(x \comp x)}} - \unit)^+ & \lleq ((x \join (x \comp x)) - \unit)^+ = ((x - \unit) \join ((x \comp x) - \unit))^+ \lleq (x - \unit)^+\plusop \\
    & = (x - \unit)^+ \lleq \cpl{\unit}
  \end{align*}
  using Theorem \ref{theorem.square}.\ref{theorem.square.square-irreflexive-part} and (12.3).
  For Theorem \ref{theorem.square}.\ref{theorem.square.disjoint-set-forest-update-square} it remains to show that $\awrite{x}{y}{\cnv{(x \comp x)}}$ is a mapping, which follows by Theorem \ref{theorem.awrite-aread}.\ref{theorem.awrite-aread.update-mapping} since $\cnv{(x \comp x)}$ is bijective using (8.1).
\end{prfof}

\begin{prfof}[of Theorems \ref{theorem.union-sets}.\ref{theorem.union-sets.plus-arc-decompose} and \ref{theorem.union-sets}.\ref{theorem.union-sets.update-acyclic-4}]
  For Theorem \ref{theorem.union-sets}.\ref{theorem.union-sets.plus-arc-decompose} assume $y$ is an arc.
  We first show $(x \join y)^* = x^* \join (x^* \comp y \comp x^*)$ \cite[Proposition 3]{BerghammerStruth2010}.
  Note that
  \begin{align*}
    (x \join y) \comp (x^* \join (x^* \comp y \comp x^*)) & = x^+ \join (x^+ \comp y \comp x^*) \join (y \comp x^*) \join (y \comp x^* \comp y \comp x^*) \\
    & = x^+ \join (x^+ \comp y \comp x^*) \join (y \comp x^*) \\
    & = x^+ \join (x^* \comp y \comp x^*)
  \end{align*}
  using (10).
  This implies $(x \join y) \comp (x^* \join (x^* \comp y \comp x^*)) \lleq x^* \join (x^* \comp y \comp x^*)$, whence $(x \join y)^* \lleq x^* \join (x^* \comp y \comp x^*)$ by Kleene star induction.
  The opposite inequality follows using (12.1).
  Hence
  \begin{align*}
    (x \join y)^+ = (x \join y) \comp (x \join y)^* = x^+ \join (x^* \comp y \comp x^*)
  \end{align*}

  For Theorem \ref{theorem.union-sets}.\ref{theorem.union-sets.update-acyclic-4} note first that assumption $y \meet (p^* \comp w) = \lbot$ is equivalent to $\cnv{y} \comp p^* \comp w \lleq \cpl{\unit}$ using (11.1) and (11.2).
  The latter is equivalent to $p^* \comp w \comp \cnv{y} \comp p^* \lleq \cpl{\unit}$ using (12.1) and (11.3).
  Second,
  \begin{align*}
    \awrite{p}{w}{y} - \unit \lleq (w \meet \cnv{y}) \join (p - \unit) = (w \comp \cnv{y}) \join (p - \unit)
  \end{align*}
  using (1.5).
  Hence
  \begin{align*}
    (\awrite{p}{w}{y} - \unit)^+ & \lleq ((w \comp \cnv{y}) \join (p - \unit))^+ \\
    & = (p - \unit)^+ \join ((p - \unit)^* \comp w \comp \cnv{y} \comp (p - \unit)^*) \\
    & \lleq (p - \unit)^+ \join (p^* \comp w \comp \cnv{y} \comp p^*) \\
    & \lleq \cpl{\unit}
  \end{align*}
  by Theorem \ref{theorem.union-sets}.\ref{theorem.union-sets.plus-arc-decompose} since $w \comp \cnv{y}$ is an arc using (1.5) and (9).
\end{prfof}

\end{document}

%% file: afp-dict.tex
\pgfkeyssetvalue{/line numbers/theorem union_sets_by_rank}{More_Disjoint_Set_Forests.thy\#L2527}
\pgfkeyssetvalue{/line numbers/theorem init_ranks}{More_Disjoint_Set_Forests.thy\#L2169}
\pgfkeyssetvalue{/line numbers/theorem find_path_splitting}{More_Disjoint_Set_Forests.thy\#L1808}
\pgfkeyssetvalue{/line numbers/theorem find_path_halving}{More_Disjoint_Set_Forests.thy\#L1173}
\pgfkeyssetvalue{/line numbers/theorem init_sets}{More_Disjoint_Set_Forests.thy\#L619}
\pgfkeyssetvalue{/line numbers/theorem union_sets_2}{Disjoint_Set_Forests.thy\#L2220}
\pgfkeyssetvalue{/line numbers/theorem union_sets}{Disjoint_Set_Forests.thy\#L2180}
\pgfkeyssetvalue{/line numbers/theorem find_set_path_compression_3}{Disjoint_Set_Forests.thy\#L1878}
\pgfkeyssetvalue{/line numbers/theorem find_set_path_compression_2}{Disjoint_Set_Forests.thy\#L1862}
\pgfkeyssetvalue{/line numbers/theorem find_set_path_compression_1}{Disjoint_Set_Forests.thy\#L1844}
\pgfkeyssetvalue{/line numbers/theorem find_set_path_compression}{Disjoint_Set_Forests.thy\#L1835}
\pgfkeyssetvalue{/line numbers/theorem path_compression}{Disjoint_Set_Forests.thy\#L1799}
\pgfkeyssetvalue{/line numbers/theorem path_compression_assign}{Disjoint_Set_Forests.thy\#L1306}
\pgfkeyssetvalue{/line numbers/theorem find_set}{Disjoint_Set_Forests.thy\#L1126}
\pgfkeyssetvalue{/line numbers/theorem make_set_2}{Disjoint_Set_Forests.thy\#L888}
\pgfkeyssetvalue{/line numbers/theorem make_set}{Disjoint_Set_Forests.thy\#L880}
\pgfkeyssetvalue{/line numbers/lemma path_compression_preserves_rank_property}{More_Disjoint_Set_Forests.thy\#L2472}
\pgfkeyssetvalue{/line numbers/lemma union_sets_1_skip}{More_Disjoint_Set_Forests.thy\#L2433}
\pgfkeyssetvalue{/line numbers/lemma union_sets_1_swap}{More_Disjoint_Set_Forests.thy\#L2256}
\pgfkeyssetvalue{/line numbers/lemma S_power_point_or_bot}{More_Disjoint_Set_Forests.thy\#L2122}
\pgfkeyssetvalue{/line numbers/lemma top_S'_sup_conv_Z}{More_Disjoint_Set_Forests.thy\#L2118}
\pgfkeyssetvalue{/line numbers/lemma Z_sup_conv_S'_top}{More_Disjoint_Set_Forests.thy\#L2114}
\pgfkeyssetvalue{/line numbers/lemma S'_star_connex}{More_Disjoint_Set_Forests.thy\#L2110}
\pgfkeyssetvalue{/line numbers/lemma S'_connected}{More_Disjoint_Set_Forests.thy\#L2106}
\pgfkeyssetvalue{/line numbers/lemma Z_below_S'_star}{More_Disjoint_Set_Forests.thy\#L2102}
\pgfkeyssetvalue{/line numbers/lemma S'_star_Z_top}{More_Disjoint_Set_Forests.thy\#L2065}
\pgfkeyssetvalue{/line numbers/lemma S'_regular}{More_Disjoint_Set_Forests.thy\#L2061}
\pgfkeyssetvalue{/line numbers/lemma M_regular}{More_Disjoint_Set_Forests.thy\#L2057}
\pgfkeyssetvalue{/line numbers/lemma S_irreflexive}{More_Disjoint_Set_Forests.thy\#L2033}
\pgfkeyssetvalue{/line numbers/lemma M_is_Z_iff_1_is_top}{More_Disjoint_Set_Forests.thy\#L2019}
\pgfkeyssetvalue{/line numbers/lemma S'_var}{More_Disjoint_Set_Forests.thy\#L2002}
\pgfkeyssetvalue{/line numbers/lemma M_point_or_bot}{More_Disjoint_Set_Forests.thy\#L1996}
\pgfkeyssetvalue{/line numbers/lemma M_bot_iff_S_not_surjective}{More_Disjoint_Set_Forests.thy\#L1980}
\pgfkeyssetvalue{/line numbers/lemma S_mapping}{More_Disjoint_Set_Forests.thy\#L1976}
\pgfkeyssetvalue{/line numbers/lemma S'_irreflexive}{More_Disjoint_Set_Forests.thy\#L1966}
\pgfkeyssetvalue{/line numbers/lemma S'_Z}{More_Disjoint_Set_Forests.thy\#L1962}
\pgfkeyssetvalue{/line numbers/lemma S'_injective}{More_Disjoint_Set_Forests.thy\#L1958}
\pgfkeyssetvalue{/line numbers/lemma S'_univalent}{More_Disjoint_Set_Forests.thy\#L1954}
\pgfkeyssetvalue{/line numbers/lemma M_point_iff_S_surjective}{More_Disjoint_Set_Forests.thy\#L1932}
\pgfkeyssetvalue{/line numbers/lemma S_inf_1_below_conv_Z}{More_Disjoint_Set_Forests.thy\#L1916}
\pgfkeyssetvalue{/line numbers/lemma S_inf_1_below_Z}{More_Disjoint_Set_Forests.thy\#L1901}
\pgfkeyssetvalue{/line numbers/lemma top_S_sup_conv_Z}{More_Disjoint_Set_Forests.thy\#L1897}
\pgfkeyssetvalue{/line numbers/lemma Z_sup_conv_S_top}{More_Disjoint_Set_Forests.thy\#L1893}
\pgfkeyssetvalue{/line numbers/lemma S_star_connex}{More_Disjoint_Set_Forests.thy\#L1889}
\pgfkeyssetvalue{/line numbers/lemma S_connected}{More_Disjoint_Set_Forests.thy\#L1885}
\pgfkeyssetvalue{/line numbers/lemma Z_below_S_star}{More_Disjoint_Set_Forests.thy\#L1876}
\pgfkeyssetvalue{/line numbers/lemma conv_Z_Z}{More_Disjoint_Set_Forests.thy\#L1872}
\pgfkeyssetvalue{/line numbers/lemma path_splitting_3}{More_Disjoint_Set_Forests.thy\#L1758}
\pgfkeyssetvalue{/line numbers/lemma path_splitting_2}{More_Disjoint_Set_Forests.thy\#L1593}
\pgfkeyssetvalue{/line numbers/lemma path_splitting_1}{More_Disjoint_Set_Forests.thy\#L1572}
\pgfkeyssetvalue{/line numbers/lemma path_splitting_invariant_aux}{More_Disjoint_Set_Forests.thy\#L1534}
\pgfkeyssetvalue{/line numbers/lemma path_splitting_invariant_aux_1}{More_Disjoint_Set_Forests.thy\#L1206}
\pgfkeyssetvalue{/line numbers/lemma path_halving_3}{More_Disjoint_Set_Forests.thy\#L1131}
\pgfkeyssetvalue{/line numbers/lemma path_halving_2}{More_Disjoint_Set_Forests.thy\#L884}
\pgfkeyssetvalue{/line numbers/lemma path_halving_1}{More_Disjoint_Set_Forests.thy\#L861}
\pgfkeyssetvalue{/line numbers/lemma path_halving_invariant_aux}{More_Disjoint_Set_Forests.thy\#L802}
\pgfkeyssetvalue{/line numbers/lemma path_halving_invariant_aux_1}{More_Disjoint_Set_Forests.thy\#L690}
\pgfkeyssetvalue{/line numbers/lemma disjoint_set_forest_update_square_point}{More_Disjoint_Set_Forests.thy\#L515}
\pgfkeyssetvalue{/line numbers/lemma disjoint_set_forest_update_square}{More_Disjoint_Set_Forests.thy\#L501}
\pgfkeyssetvalue{/line numbers/lemma acyclic_update_square}{More_Disjoint_Set_Forests.thy\#L485}
\pgfkeyssetvalue{/line numbers/lemma square_irreflexive_part_2}{More_Disjoint_Set_Forests.thy\#L481}
\pgfkeyssetvalue{/line numbers/lemma square_irreflexive_part}{More_Disjoint_Set_Forests.thy\#L460}
\pgfkeyssetvalue{/line numbers/lemma star_irreflexive_part}{More_Disjoint_Set_Forests.thy\#L456}
\pgfkeyssetvalue{/line numbers/lemma star_irreflexive_part_eq}{More_Disjoint_Set_Forests.thy\#L452}
\pgfkeyssetvalue{/line numbers/lemma acyclic_plus_loop}{More_Disjoint_Set_Forests.thy\#L424}
\pgfkeyssetvalue{/line numbers/lemma fc_update_square}{More_Disjoint_Set_Forests.thy\#L374}
\pgfkeyssetvalue{/line numbers/lemma diagonal_update_square}{More_Disjoint_Set_Forests.thy\#L354}
\pgfkeyssetvalue{/line numbers/lemma diagonal_update_square_aux}{More_Disjoint_Set_Forests.thy\#L335}
\pgfkeyssetvalue{/line numbers/lemma acyclic_square}{More_Disjoint_Set_Forests.thy\#L311}
\pgfkeyssetvalue{/line numbers/lemma update_square_ub_plus}{More_Disjoint_Set_Forests.thy\#L307}
\pgfkeyssetvalue{/line numbers/lemma update_square_plus}{More_Disjoint_Set_Forests.thy\#L303}
\pgfkeyssetvalue{/line numbers/lemma even_odd_root}{More_Disjoint_Set_Forests.thy\#L219}
\pgfkeyssetvalue{/line numbers/lemma omit_redundant_points_3}{More_Disjoint_Set_Forests.thy\#L214}
\pgfkeyssetvalue{/line numbers/lemma omit_redundant_points_2}{More_Disjoint_Set_Forests.thy\#L180}
\pgfkeyssetvalue{/line numbers/lemma update_mapping_swap}{More_Disjoint_Set_Forests.thy\#L166}
\pgfkeyssetvalue{/line numbers/lemma update_univalent_swap}{More_Disjoint_Set_Forests.thy\#L157}
\pgfkeyssetvalue{/line numbers/lemma update_injective_swap_2}{More_Disjoint_Set_Forests.thy\#L152}
\pgfkeyssetvalue{/line numbers/lemma update_injective_swap}{More_Disjoint_Set_Forests.thy\#L87}
\pgfkeyssetvalue{/line numbers/lemma update_split}{More_Disjoint_Set_Forests.thy\#L82}
\pgfkeyssetvalue{/line numbers/lemma update_same_3}{More_Disjoint_Set_Forests.thy\#L76}
\pgfkeyssetvalue{/line numbers/lemma update_same}{More_Disjoint_Set_Forests.thy\#L62}
\pgfkeyssetvalue{/line numbers/lemma update_top}{More_Disjoint_Set_Forests.thy\#L58}
\pgfkeyssetvalue{/line numbers/lemma update_bot}{More_Disjoint_Set_Forests.thy\#L54}
\pgfkeyssetvalue{/line numbers/lemma update_point_get}{More_Disjoint_Set_Forests.thy\#L50}
\pgfkeyssetvalue{/line numbers/lemma update_same_sub}{More_Disjoint_Set_Forests.thy\#L43}
\pgfkeyssetvalue{/line numbers/lemma update_square_ub}{More_Disjoint_Set_Forests.thy\#L39}
\pgfkeyssetvalue{/line numbers/lemma update_ub}{More_Disjoint_Set_Forests.thy\#L35}
\pgfkeyssetvalue{/line numbers/lemma update_square}{More_Disjoint_Set_Forests.thy\#L21}
\pgfkeyssetvalue{/line numbers/lemma union_sets_function}{Disjoint_Set_Forests.thy\#L2214}
\pgfkeyssetvalue{/line numbers/lemma union_sets_exists}{Disjoint_Set_Forests.thy\#L2208}
\pgfkeyssetvalue{/line numbers/lemma union_sets_1}{Disjoint_Set_Forests.thy\#L2003}
\pgfkeyssetvalue{/line numbers/lemma find_set_path_compression_find_set_pathcompression}{Disjoint_Set_Forests.thy\#L1972}
\pgfkeyssetvalue{/line numbers/lemma find_set_path_compression_path_compression_semantics}{Disjoint_Set_Forests.thy\#L1949}
\pgfkeyssetvalue{/line numbers/lemma find_set_path_compression_find_set}{Disjoint_Set_Forests.thy\#L1930}
\pgfkeyssetvalue{/line numbers/lemma find_set_path_compression_function}{Disjoint_Set_Forests.thy\#L1913}
\pgfkeyssetvalue{/line numbers/lemma find_set_path_compression_exists}{Disjoint_Set_Forests.thy\#L1907}
\pgfkeyssetvalue{/line numbers/lemma path_compression_function}{Disjoint_Set_Forests.thy\#L1822}
\pgfkeyssetvalue{/line numbers/lemma path_compression_exists}{Disjoint_Set_Forests.thy\#L1816}
\pgfkeyssetvalue{/line numbers/lemma path_compression_3}{Disjoint_Set_Forests.thy\#L1795}
\pgfkeyssetvalue{/line numbers/lemma path_compression_3a}{Disjoint_Set_Forests.thy\#L1750}
\pgfkeyssetvalue{/line numbers/lemma path_compression_2}{Disjoint_Set_Forests.thy\#L1341}
\pgfkeyssetvalue{/line numbers/lemma path_compression_1}{Disjoint_Set_Forests.thy\#L1337}
\pgfkeyssetvalue{/line numbers/lemma path_compression_1b}{Disjoint_Set_Forests.thy\#L1333}
\pgfkeyssetvalue{/line numbers/lemma path_compression_1a}{Disjoint_Set_Forests.thy\#L1326}
\pgfkeyssetvalue{/line numbers/lemma update_acyclic_6}{Disjoint_Set_Forests.thy\#L1300}
\pgfkeyssetvalue{/line numbers/lemma path_compression_exact}{Disjoint_Set_Forests.thy\#L1191}
\pgfkeyssetvalue{/line numbers/lemma find_set_function}{Disjoint_Set_Forests.thy\#L1159}
\pgfkeyssetvalue{/line numbers/lemma root_point}{Disjoint_Set_Forests.thy\#L1153}
\pgfkeyssetvalue{/line numbers/lemma find_set_exists}{Disjoint_Set_Forests.thy\#L1141}
\pgfkeyssetvalue{/line numbers/lemma find_set_3}{Disjoint_Set_Forests.thy\#L1073}
\pgfkeyssetvalue{/line numbers/lemma find_set_2}{Disjoint_Set_Forests.thy\#L1014}
\pgfkeyssetvalue{/line numbers/lemma find_set_1}{Disjoint_Set_Forests.thy\#L1009}
\pgfkeyssetvalue{/line numbers/lemma forest_mutually_reachable_2}{Disjoint_Set_Forests.thy\#L988}
\pgfkeyssetvalue{/line numbers/lemma forest_mutually_reachable}{Disjoint_Set_Forests.thy\#L953}
\pgfkeyssetvalue{/line numbers/lemma make_set_function}{Disjoint_Set_Forests.thy\#L916}
\pgfkeyssetvalue{/line numbers/lemma make_set_exists}{Disjoint_Set_Forests.thy\#L910}
\pgfkeyssetvalue{/line numbers/lemma put_put_different}{Disjoint_Set_Forests.thy\#L833}
\pgfkeyssetvalue{/line numbers/lemma put_put_different_vector}{Disjoint_Set_Forests.thy\#L815}
\pgfkeyssetvalue{/line numbers/lemma put_get_different}{Disjoint_Set_Forests.thy\#L803}
\pgfkeyssetvalue{/line numbers/lemma put_get_different_vector}{Disjoint_Set_Forests.thy\#L785}
\pgfkeyssetvalue{/line numbers/lemma path_compression_invariant_simplify}{Disjoint_Set_Forests.thy\#L759}
\pgfkeyssetvalue{/line numbers/lemma loop_root_2}{Disjoint_Set_Forests.thy\#L728}
\pgfkeyssetvalue{/line numbers/lemma root_root}{Disjoint_Set_Forests.thy\#L724}
\pgfkeyssetvalue{/line numbers/lemma one_loop}{Disjoint_Set_Forests.thy\#L701}
\pgfkeyssetvalue{/line numbers/lemma loop_root}{Disjoint_Set_Forests.thy\#L685}
\pgfkeyssetvalue{/line numbers/lemma same_root}{Disjoint_Set_Forests.thy\#L677}
\pgfkeyssetvalue{/line numbers/lemma same_roots}{Disjoint_Set_Forests.thy\#L669}
\pgfkeyssetvalue{/line numbers/lemma same_roots_sub}{Disjoint_Set_Forests.thy\#L651}
\pgfkeyssetvalue{/line numbers/lemma same_component_same_root}{Disjoint_Set_Forests.thy\#L636}
\pgfkeyssetvalue{/line numbers/lemma same_component_same_root_sub}{Disjoint_Set_Forests.thy\#L624}
\pgfkeyssetvalue{/line numbers/lemma univalent_root_successors}{Disjoint_Set_Forests.thy\#L612}
\pgfkeyssetvalue{/line numbers/lemma root_same_component_vector}{Disjoint_Set_Forests.thy\#L608}
\pgfkeyssetvalue{/line numbers/lemma root_vector_inf}{Disjoint_Set_Forests.thy\#L604}
\pgfkeyssetvalue{/line numbers/lemma root_vector}{Disjoint_Set_Forests.thy\#L600}
\pgfkeyssetvalue{/line numbers/lemma root_same_component}{Disjoint_Set_Forests.thy\#L596}
\pgfkeyssetvalue{/line numbers/lemma roots_transitive_successor_loop}{Disjoint_Set_Forests.thy\#L590}
\pgfkeyssetvalue{/line numbers/lemma roots_successor_loop}{Disjoint_Set_Forests.thy\#L586}
\pgfkeyssetvalue{/line numbers/lemma root_transitive_successor_loop}{Disjoint_Set_Forests.thy\#L582}
\pgfkeyssetvalue{/line numbers/lemma root_successor_loop}{Disjoint_Set_Forests.thy\#L578}
\pgfkeyssetvalue{/line numbers/lemma root_var}{Disjoint_Set_Forests.thy\#L574}
\pgfkeyssetvalue{/line numbers/lemma update_acyclic_5}{Disjoint_Set_Forests.thy\#L555}
\pgfkeyssetvalue{/line numbers/lemma update_acyclic_4}{Disjoint_Set_Forests.thy\#L527}
\pgfkeyssetvalue{/line numbers/lemma plus_arc_decompose}{Disjoint_Set_Forests.thy\#L523}
\pgfkeyssetvalue{/line numbers/lemma plus_rectangle_decompose}{Disjoint_Set_Forests.thy\#L507}
\pgfkeyssetvalue{/line numbers/lemma star_arc_decompose}{Disjoint_Set_Forests.thy\#L503}
\pgfkeyssetvalue{/line numbers/lemma star_rectangle_decompose}{Disjoint_Set_Forests.thy\#L482}
\pgfkeyssetvalue{/line numbers/lemma arc_star_arc}{Disjoint_Set_Forests.thy\#L478}
\pgfkeyssetvalue{/line numbers/lemma rectangle_star_rectangle}{Disjoint_Set_Forests.thy\#L474}
\pgfkeyssetvalue{/line numbers/lemma update_acyclic_3}{Disjoint_Set_Forests.thy\#L466}
\pgfkeyssetvalue{/line numbers/lemma update_acyclic_2}{Disjoint_Set_Forests.thy\#L447}
\pgfkeyssetvalue{/line numbers/lemma update_acyclic_1}{Disjoint_Set_Forests.thy\#L413}
\pgfkeyssetvalue{/line numbers/lemma fc_via_root}{Disjoint_Set_Forests.thy\#L399}
\pgfkeyssetvalue{/line numbers/lemma fc_wcc}{Disjoint_Set_Forests.thy\#L395}
\pgfkeyssetvalue{/line numbers/lemma fc_top}{Disjoint_Set_Forests.thy\#L391}
\pgfkeyssetvalue{/line numbers/lemma fc_one}{Disjoint_Set_Forests.thy\#L387}
\pgfkeyssetvalue{/line numbers/lemma fc_bot}{Disjoint_Set_Forests.thy\#L383}
\pgfkeyssetvalue{/line numbers/lemma fc_plus}{Disjoint_Set_Forests.thy\#L379}
\pgfkeyssetvalue{/line numbers/lemma fc_star}{Disjoint_Set_Forests.thy\#L375}
\pgfkeyssetvalue{/line numbers/lemma fc_idempotent}{Disjoint_Set_Forests.thy\#L371}
\pgfkeyssetvalue{/line numbers/lemma fc_isotone}{Disjoint_Set_Forests.thy\#L367}
\pgfkeyssetvalue{/line numbers/lemma fc_increasing}{Disjoint_Set_Forests.thy\#L363}
\pgfkeyssetvalue{/line numbers/lemma fc_equivalence}{Disjoint_Set_Forests.thy\#L355}
\pgfkeyssetvalue{/line numbers/lemma wcc_sup_wcc}{Disjoint_Set_Forests.thy\#L347}
\pgfkeyssetvalue{/line numbers/lemma forest_components_wcc}{Disjoint_Set_Forests.thy\#L343}
\pgfkeyssetvalue{/line numbers/lemma wcc_without_loops}{Disjoint_Set_Forests.thy\#L339}
\pgfkeyssetvalue{/line numbers/lemma wcc_with_loops}{Disjoint_Set_Forests.thy\#L335}
\pgfkeyssetvalue{/line numbers/lemma wcc_top}{Disjoint_Set_Forests.thy\#L331}
\pgfkeyssetvalue{/line numbers/lemma wcc_one}{Disjoint_Set_Forests.thy\#L327}
\pgfkeyssetvalue{/line numbers/lemma wcc_bot}{Disjoint_Set_Forests.thy\#L323}
\pgfkeyssetvalue{/line numbers/lemma wcc_galois}{Disjoint_Set_Forests.thy\#L319}
\pgfkeyssetvalue{/line numbers/lemma wcc_below_wcc}{Disjoint_Set_Forests.thy\#L315}
\pgfkeyssetvalue{/line numbers/lemma wcc_idempotent}{Disjoint_Set_Forests.thy\#L311}
\pgfkeyssetvalue{/line numbers/lemma wcc_isotone}{Disjoint_Set_Forests.thy\#L307}
\pgfkeyssetvalue{/line numbers/lemma wcc_increasing}{Disjoint_Set_Forests.thy\#L303}
\pgfkeyssetvalue{/line numbers/lemma wcc_equivalence}{Disjoint_Set_Forests.thy\#L295}
\pgfkeyssetvalue{/line numbers/lemma omit_redundant_points}{Disjoint_Set_Forests.thy\#L254}
\pgfkeyssetvalue{/line numbers/lemma update_inf_different}{Disjoint_Set_Forests.thy\#L234}
\pgfkeyssetvalue{/line numbers/lemma update_inf_same}{Disjoint_Set_Forests.thy\#L230}
\pgfkeyssetvalue{/line numbers/lemma update_inf}{Disjoint_Set_Forests.thy\#L226}
\pgfkeyssetvalue{/line numbers/lemma get_put}{Disjoint_Set_Forests.thy\#L189}
\pgfkeyssetvalue{/line numbers/lemma put_put}{Disjoint_Set_Forests.thy\#L185}
\pgfkeyssetvalue{/line numbers/lemma put_get}{Disjoint_Set_Forests.thy\#L180}
\pgfkeyssetvalue{/line numbers/lemma put_get_sub}{Disjoint_Set_Forests.thy\#L157}
\pgfkeyssetvalue{/line numbers/lemma update_postcondition}{Disjoint_Set_Forests.thy\#L142}
\pgfkeyssetvalue{/line numbers/lemma read_point}{Disjoint_Set_Forests.thy\#L136}
\pgfkeyssetvalue{/line numbers/lemma read_bijective}{Disjoint_Set_Forests.thy\#L130}
\pgfkeyssetvalue{/line numbers/lemma read_surjective}{Disjoint_Set_Forests.thy\#L124}
\pgfkeyssetvalue{/line numbers/lemma read_injective}{Disjoint_Set_Forests.thy\#L118}
\pgfkeyssetvalue{/line numbers/lemma update_mapping}{Disjoint_Set_Forests.thy\#L110}
\pgfkeyssetvalue{/line numbers/lemma update_total}{Disjoint_Set_Forests.thy\#L93}
\pgfkeyssetvalue{/line numbers/lemma update_univalent}{Disjoint_Set_Forests.thy\#L61}
\pgfkeyssetvalue{/line numbers/lemma arc_rectangle}{Disjoint_Set_Forests.thy\#L42}
\pgfkeyssetvalue{/line numbers/lemma distinct_points}{Disjoint_Set_Forests.thy\#L781}
\pgfkeyssetvalue{/line numbers/lemma mapping_inf_point_arc}{More_Disjoint_Set_Forests.thy\#L173}